\documentclass[aps,prl,superscriptaddress,reprint]{revtex4-1}
\usepackage{amsmath}
\usepackage{amssymb}
\usepackage{graphicx}
\usepackage[colorlinks,urlcolor=blue]{hyperref}
\usepackage{url}
\usepackage{color,xcolor}
\usepackage{ulem}
\usepackage{float}
\usepackage{epstopdf}
\begin{document}

\title{Strongly anisotropic electronic and magnetic structures in \\
oxide dichlorides RuOCl$_2$ and OsOCl$_2$}
\author{Yang Zhang}
\author{Ling-Fang Lin}
\affiliation{Department of Physics and Astronomy, University of Tennessee, Knoxville, TN 37996, USA}
\author{Adriana Moreo}
\affiliation{Department of Physics and Astronomy, University of Tennessee, Knoxville, TN 37996, USA}
\affiliation{Materials Science and Technology Division, Oak Ridge National Laboratory, Oak Ridge, TN 37831, USA}
\author{Thomas A. Maier}
\affiliation{Computational Sciences and Engineering Division, Oak Ridge National Laboratory, Oak Ridge, Tennessee 37831, USA}
\author{Gonzalo Alvarez}
\affiliation{Computational Sciences \& Engineering Division and Center for Nanophase Materials Sciences, Oak Ridge National Laboratory, Oak Ridge, TN 37831, USA}
\author{Elbio Dagotto}
\affiliation{Department of Physics and Astronomy, University of Tennessee, Knoxville, TN 37996, USA}
\affiliation{Materials Science and Technology Division, Oak Ridge National Laboratory, Oak Ridge, TN 37831, USA}

\date{\today}

\begin{abstract}
The van der Waals oxide dichlorides $M$O$X_2$ ($M$ = V, Ta, Nb, Ru, and Os; $X$ = halogen element), with different electronic densities, are attracting considerable attention. Ferroelectricity, spin-singlet formation,
and orbital-selective Peierls phases were reported in this family with $d^1$ or $d^2$ electronic configurations, all believed to be caused by the strongly anisotropic electronic orbital degree of freedom. Here, using density
functional theory and density matrix renormalization group methods, we
investigate the electronic and magnetic properties of RuOCl$_2$ and OsOCl$_2$ with $d^4$ electronic configurations. Different from a
previous study using VOI$_2$ with $d^1$ configuration, these systems with $4d^4$ or $5d^4$ do not exhibit a ferroelectric instability along the
$a$-axis. Due to the fully-occupied $d_{xy}$ orbital in RuOCl$_2$ and OsOCl$_2$, the Peierls instability distortion disappears
along the $b$-axis, leading to an undistorted I${\rm mmm}$ phase (No. 71). Furthermore, we observe
strongly anisotropic electronic and magnetic structures along the $a$-axis. For this reason, the materials of our focus can be regarded as ``effective 1D'' systems even when they apparently have a dominant two-dimensional lattice geometry.
The large crystal-field splitting energy (between $d_{xz/yz}$ and $d_{xy}$ orbitals) and large hopping between nearest-neighbor
Ru and Os atoms suppresses the $J = 0$ singlet state in $M$OCl$_2$ ($M$ = Ru or Os) with
electronic density $n = 4$, resulting in a spin-1 system.
Moreover, we find staggered antiferromagnetic order with $\pi$ wavevector along the $M$-O chain direction ($a$-axis) while the
magnetic coupling along the $b$-axis is weak. Based on Wannier functions from first-principles calculations, we calculated
the relevant hopping amplitudes and crystal-field splitting energies of the $t_{2g}$ orbitals for the Os atoms to construct a multi-orbital Hubbard model for the $M$-O chains. Staggered AFM
with $\uparrow$-$\downarrow$-$\uparrow$-$\downarrow$ spin structure dominates in our DMRG calculations, in agreement with DFT calculations.
Our results for RuOCl$_2$ and OsOCl$_2$ provide guidance to experimentalists
and theorists working on this interesting family of oxide dichlorides.
\end{abstract}

\maketitle
\section{I. Introduction}
One-dimensional (1D) material systems continue to attract considerable attention due to their rich physical properties induced by their 1D geometry and reduced dimensional phase space~\cite{Dagotto:rmp94,Grioni:JPCM,Monceau:ap,Dagotto:Rmp,Bertini:Rmp,Gangadharaiah:prl,Herbrych:pnas,Herbrych:nc21,Lin:prl21}.  In these systems, many interesting phenomena have been found that are driven by intertwined charge, spin, orbital, and lattice degrees of freedom. For example, driven by electronic correlation effects (i.e. Hubbard repulsion $U$ and Hund coupling $J_H$), high critical temperature superconductivity was reported in 1D copper or iron chains and ladders~\cite{cu-ladder1,cu-ladder2,cu-ladder3,Takahashi:Nm,Ying:prb17,Zhang:prb17,Zhang:prb18,Zhang:prb19}. By considering the phonon instability caused by the coupling between empty $d$ and fully occupied O $2p$ states, ferroelectricity was found in the chain compound WO$X_4$ ($X$ = halogen element)~\cite{Lin:prm}. Furthermore, by mixing spin-phonon and charge-phonon instabilities, multiferroelectric states were predicted in some 1D systems~\cite{Brink:jpcm,Lin:prm17,Dong:nsr,Zhang:prb20-2}. Due to the partial or complete condensation of excitations, a charge density wave or a spin density wave were also reported in some 1D systems~\cite{Grioni:JPCM,Wang:prb,Gooth:nature,Zhang:prbcdw}.

\begin{figure}
\centering
\includegraphics[width=0.48\textwidth]{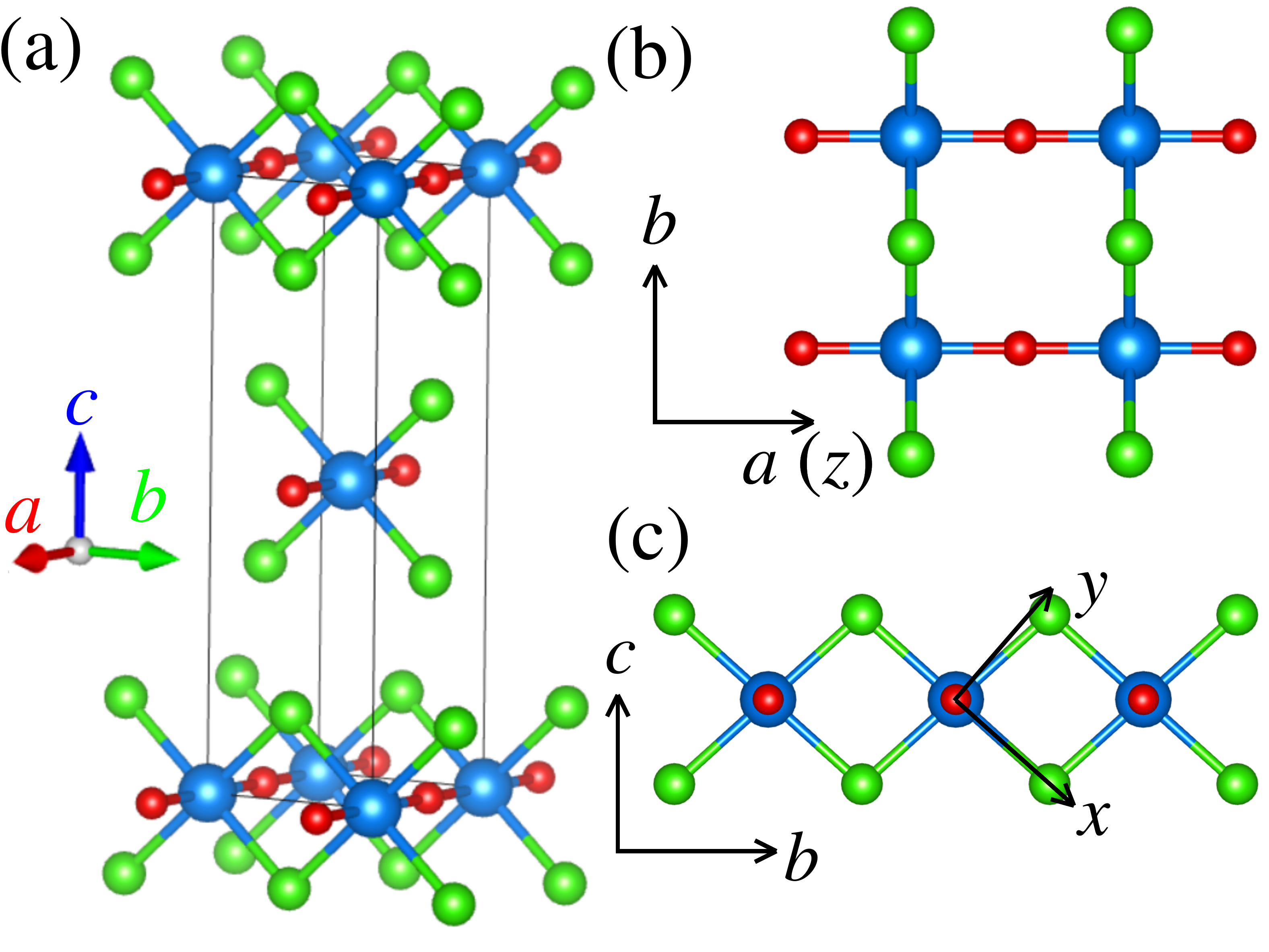}
\caption{ (a-c) Schematic crystal structure of the undistorted parent phase of $M$O$X_2$ ($M$ = V, Ta, Nb, Ru and Os; $X$ = halogen element): Blue = $M$ ($M$ = V, Ta, Nb, Ru and Os); red = O; green = $X$ ($X$ = halogen element). (a) Conventional cell of the bulk structure. (b) Sketch of the $ab$ plane. (c) $MX_2$ chain along the $b$-axis. Note that the local $z$-axis is the $a$-axis, while the local $x$- or $y$-axis is along the $M$-$X$ bond directions, leading to $d_{xy}$ orbitals lying on the $bc$ plane.}
\label{Fig1}
\end{figure}

A wide variety of real materials also have dominant 1D-like physical properties, even without restrictive 1D geometries in their crystal structure, due to the strongly anisotropic electronic orbital degree of freedom. Recently, several different interesting 1D physical properties were reported in oxide dichlorides $M$O$X_2$ ($M$ = V, Ta, Nb, Ru and Os; $X$ = halogen element) with various electronic densities $n$ for the $M$ atoms~\cite{Hillebrecht:jac,Schnering:ac,Ruck:acc,Tan:prb,Xu:prl,Jia:nh,Wang:prm20}. The parent phase of $M$O$X_2$ ($M$ = V, Ta, Nb, Os; $X$ = halogen element) is a typical member of the layered van der Waals (vdW) family~\cite{Hillebrecht:jac,Schnering:ac,Ruck:acc}, where the $M$O$_2$$X_4$ octahedra are corner-sharing along the $a$-axis, while edge-sharing along the $b$-axis [see Fig.~\ref{Fig1}]. The remarkable effective 1D-like behavior of this family can be understood from
the strong anisotropic behavior of different orbitals.

In $M$O$X_2$, the $M$O$_2$Cl$_4$ octahedra are edge-sharing connected, opening the possibility of strong overlap of $d_{xy}$ orbitals along the $xy$ plane. Due to the 1D $M$-$X$ chain geometric structures along the $b$-axis [see Figs.~\ref{Fig1}(b) and (c)], the bandstructure resulting from the $xy$ orbital displays strong anisotropy. Furthermore, the $d_{xz}$ and $d_{yz}$ orbitals also display anisotropy along the $a$-axis due to the $M$-O geometric chain, while it forms the vdW layer along the $c$-axis. For the $n = 1$ case with $d^1$ electronic configuration (V or Nb),
a ferroelectric (FE) distortion~\cite{Tan:prb,Xu:prl,Jia:nh,Ding:prb} was theoretically predicted along the $M$-O direction ($a$-axis), where the ``pseudo-Jahn-Teller'' effect caused by the coupling between empty $d$ ($d_{xz/yz}$ and $d_{3z^2-r^2}$) and O $2p$ orbitals plays an important role to stabilize the FE distortion~\cite{Zhang:prb21}. In addition, Peierls distortions were found along the $b$-axis~\cite{Jia:nh,Zhang:prb21}, resulting in a spin-singlet configuration for the $d_{xy}$ orbitals~\cite{Zhang:prb21}, due to the formation of molecular states in the $d_{xy}$ bands. Very recently, the FE and Peierls distortions were confirmed experimentally for NbOI$_2$ with $4d^1$ electronic configuration~\cite{Fang:am21}. With additional spin-orbit coupling (SOC), a spin texture was also found at the $Y$ point along the $M$-$X$ chain direction~\cite{Ye:cpl}.

For the case of a $d^2$ electronic configurations, MoOCl$_2$ was experimentally reported to be a strongly correlated dimerized metal based on temperature-dependent transport measurements along the $M$-$X$ chain direction~\cite{Wang:prm20}. The metallic conductivity arises from the strongly anisotropic Mo-$d_{xz/yz}$ bands~\cite{Zhao:prb20,Zhang:ossp}. Furthermore, an interesting orbital-selective Peierls phase was also found to be stable in
MoOCl$_2$~\cite{Zhang:ossp}, because the intra-hopping amplitude $t$ is larger than the typical Hund couplings. This phase resembles the previously discussed orbital-selective Mott phase~\cite{OSMP,Patel:osmp,Herbrych:osmp1,zhang:2021,Lin:arxiv} but with the localized band induced by Peierls distortions instead of Hubbard interactions~\cite{Kimber:prb14,Zhang:ossp}. Moreover, highly anisotropic plasmons were discussed in the monolayer MoOCl$_2$ ~\cite{Gao:2021}.

{\it Yet almost no research has been done for other electronic densities $n$ of $M$ atoms in this family.} RuOCl$_2$ and OsOCl$_2$ with $d^4$ electronic configuration were synthesized~\cite{Hillebrecht:jac}, and it was reported that orthorhombic structures are formed with the space group I${\rm mmm}$ (No. 71) [see Fig.~\ref{Fig1}]. Due to a reduced $J_H$, often $4d/5d$ atoms favor the total $S=1$ configuration in compounds with more than half-filled $t_{2g}$ states, leading to four electrons occupying three $t_{2g}$ orbitals in both RuOCl$_2$ and OsOCl$_2$. Considering the development of different 1D behaviors caused by different $t_{2g}$ orbitals, a simple question naturally arises: Can RuOCl$_2$ and OsOCl$_2$ also display similar physical properties? In addition, with additional SOC, $d^4$ materials are expected to be nonmagnetic insulators formed by local two-hole $J = 0$ singlets~\cite{Khaliullin:prl13,Meetei:prb15}. Is it possible to obtain $J = 0$ singlets in RuOCl$_2$ and OsOCl$_2$ as well?

To answer these questions, we employ both density functional theory (DFT) and density matrix renormalization group (DMRG) methods to numerically investigate RuOCl$_2$ and OsOCl$_2$ in detail. Based on DFT calculations, we have found that there are no FE distortions and Peierls instabilities occurring along the $a$- or $b$-axis in the undistorted phase of this system. Furthermore, we also observed a strongly anisotropic electronic structure along the $a$-axis. Because of the large crystal-field splitting energy (between $d_{xz/yz}$ and $d_{xy}$ orbitals) and the large nearest-neighbor (NN) hopping, the $J = 0$ singlet ground state is suppressed in this system with the $d^4$ electronic configuration, leading to a spin-1 system. In addition, based on DFT calculations, we also found a strongly anisotropic electronic structure, with strong couplings along the $a$-axis and much weaker coupling along the $b$-axis for both RuOCl$_2$ and OsOCl$_2$. For this reason, surprisingly, these systems can be regarded as ``effective 1D'' materials, although naively they should be planar 2D systems. Using Wannier functions from first-principles calculations, we obtained the relevant hopping amplitudes and crystal-field splitting energies for the $t_{2g}$ orbitals of the Ru/Os atoms. We found that staggered spin order is the most likely magnetic ground state, with a $\pi$ wavevector order along the chain direction. Finally, we constructed a multi-orbital Hubbard model for the $M$-O chains and analyzed this model using DMRG. Our results show that staggered AFM order with $\uparrow$-$\downarrow$-$\uparrow$-$\downarrow$ spin structure is dominant, consistent with the DFT calculations.

\section{II. Method and DFT Calculations}

In the present study, first-principles DFT calculations were performed using the Vienna {\it ab initio} simulation package (VASP) code~\cite{Kresse:Prb,Kresse:Prb96,Blochl:Prb} with the projector augmented wave (PAW) method. Electronic correlations were considered by using the generalized gradient approximation (GGA) with the Perdew-Burke-Ernzerhof (PBE) potential~\cite{Perdew:Prl}. The $k$-point mesh adopted was $16\times16\times5$ for the conventional cell of the bulk system, while the plane-wave cutoff energy was $600$~eV. We have tested explicitly that this $k$-point mesh already leads to converged energies. Furthermore, both the lattice constants and atomic positions were fully relaxed until the Hellman-Feynman force on each atom was smaller than $0.01$ eV/{\AA}. The van der Waals (vdW) interactions Becke-Jonson damping vdW-D3~\cite{Grimme:jcc} were considered to deal with interactions between different layers. The phonon spectra were calculated using the density functional perturbation theory approach~\cite{Baroni:Prl,DFPT} and analyzed by the PHONONPY software~\cite{Chaput:prb,Togo:sm}. In addition to the standard DFT calculation discussed thus far, the maximally localized Wannier functions (MLWFs) method was employed to fit the Ru $4d$'s or Os $5d$'s three $t_{2g}$ bands near the Fermi level using the WANNIER90 packages~\cite{Mostofi:cpc}. All the crystal structures were visualized with the VESTA code~\cite{Momma:vesta}.

To better understand the magnetic properties, we also relaxed the crystal structures for selected different spin configurations based on the $2\times2\times1$ supercell. Furthermore, the on-site Coulomb interactions were considered by using the local spin density approximation (LSDA) plus $U$ with the Liechtenstein formulation for the double-counting term~\cite{Liechtenstein:prb}. Based on previous experimental and theoretical studies for $4d$ and $5d$ compounds~\cite{Yuan:prb17,prb11,Du:prb12}, the on-site Coulomb interaction $U$ and on-site exchange interaction $J$ were chosen as
$U = 3$~eV and $J = 0.6$~eV for RuOCl$_2$, and $U = 2$~eV and $J = 0.4$~eV for OsOCl$_2$, respectively.

\section{III. DFT Results}

\subsection{A. Structural properties}
Based on our structural optimization calculation of the bulk nonmagnetic (NM) state, the optimized crystal lattices are $a = 3.666$, $b = 3.554$ ~and $c = 11.266$ \AA~for RuOCl$_2$,  close to experimental values ($a = 3.673$, $b = 3.520$ ~and $c = 11.258$ \AA)~\cite{Hillebrecht:jac}. We also obtained the lattice constants of OsOCl$_2$ ($a = 3.718$, $b = 3.615$ ~and $c = 11.079$ \AA), also in agreement with experiments ($a = 3.701$, $b = 3.575$ ~and $c = 11.083$ \AA)~\cite{Hillebrecht:jac}.

\begin{figure}
\centering
\includegraphics[width=0.48\textwidth]{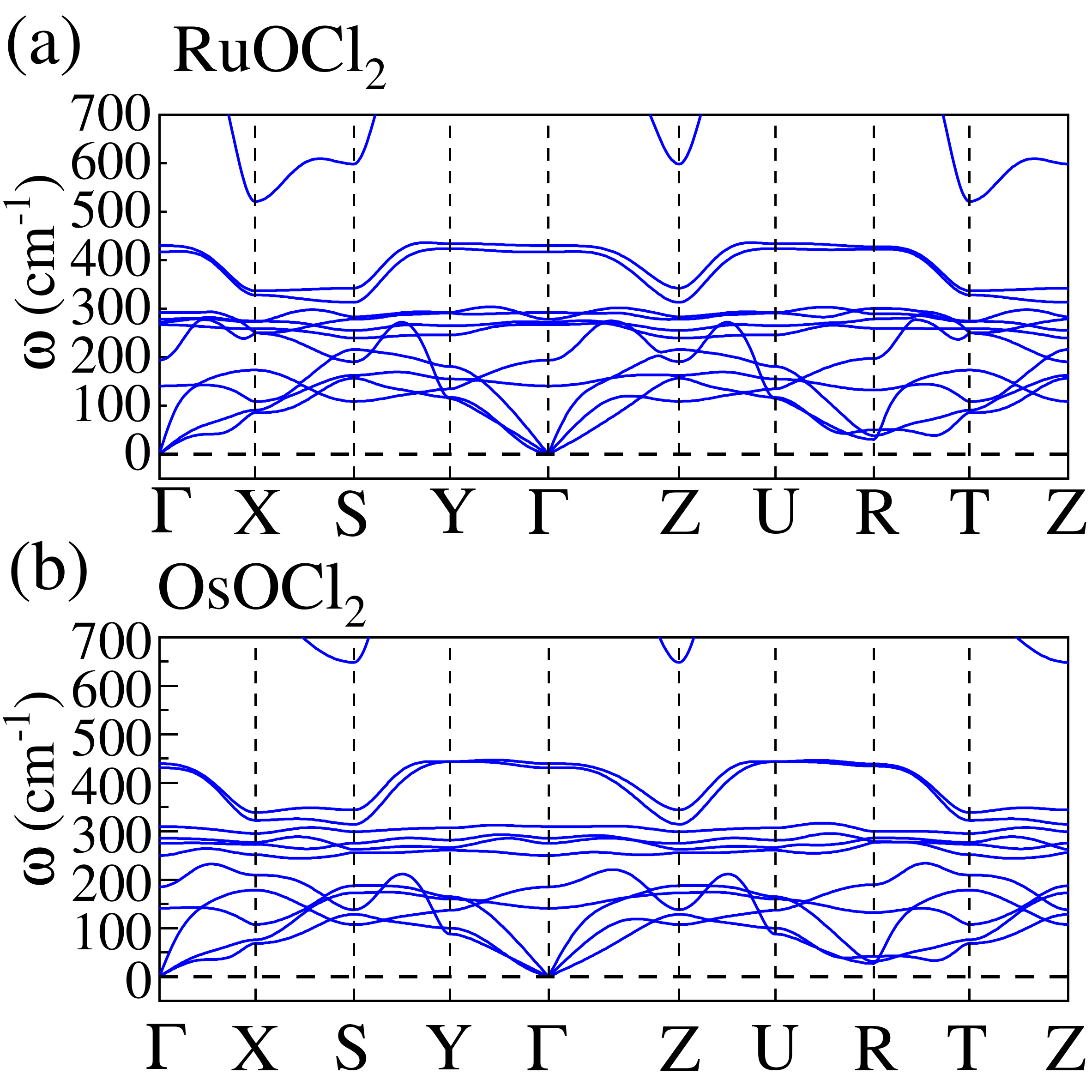}
\caption{Phonon spectrum of the bulk system for (a) RuOCl$_2$ and (b) OsOCl$_2$. A $4\times4\times1$ supercell was used in our calculations and
the nonmagnetic state. The coordinates of the high-symmetry points in the bulk Brillouin zone (BZ) are
$\Gamma$ = (0, 0, 0), X = (0.5, 0, 0), S = (0.5, 0.5, 0), Y = (0, 0.5, 0), Z = (0, 0, 0.5), U = (0.5, 0, 0.5), R = (0.5, 0.5, 0.5), and T = (0, 0.5, 0.5) in units of $2\pi$.}
\label{Fig2}
\end{figure}

Before turning to the physical properties of $M$OCl$_2$ ($M$ = Ru or Os), we discuss their structural properties. We carried out phononic dispersion calculations using a $4\times4\times1$ supercell to understand the structural stability of the undistorted I${\rm mmm}$ phase (No. 71). Figure~\ref{Fig2} indicates that there is no imaginary frequency mode obtained in the phononic dispersion spectrum for the I${\rm mmm}$ phase of RuOCl$_2$ and OsOCl$_2$. Therefore, the undistorted I${\rm mmm}$ phase of bulk $M$OCl$_2$ ($M$ = Ru or Os) is dynamically stable, in agreement with experiments~\cite{Hillebrecht:jac}.

In previous studies for this vdW family of layered oxide dichlorides $M$O$X_2$ ($M$ = V, Ta, Nb, Os; $X$ = halogen element), the ferroelectric distortion along the $a$-axis and the $M-M$ dimerization along the $b$-axis induced structural instabilities, leading to lower symmetry structures~\cite{Zhang:prb21}, such as in the case of the FE distortion in VOI$_2$ (3$d^1$)~\cite{Tan:prb,Ding:prb} or NbOI$_2$ (4$d^1$)~\cite{Jia:nh,Fang:am21} and the Peierls distortion in MoOCl$_2$ (4$d^2$)~\cite{Wang:prm20} or TaOI$_2$ (5$d^1$)~\cite{Ruck:acc}.

\begin{figure}
\centering
\includegraphics[width=0.48\textwidth]{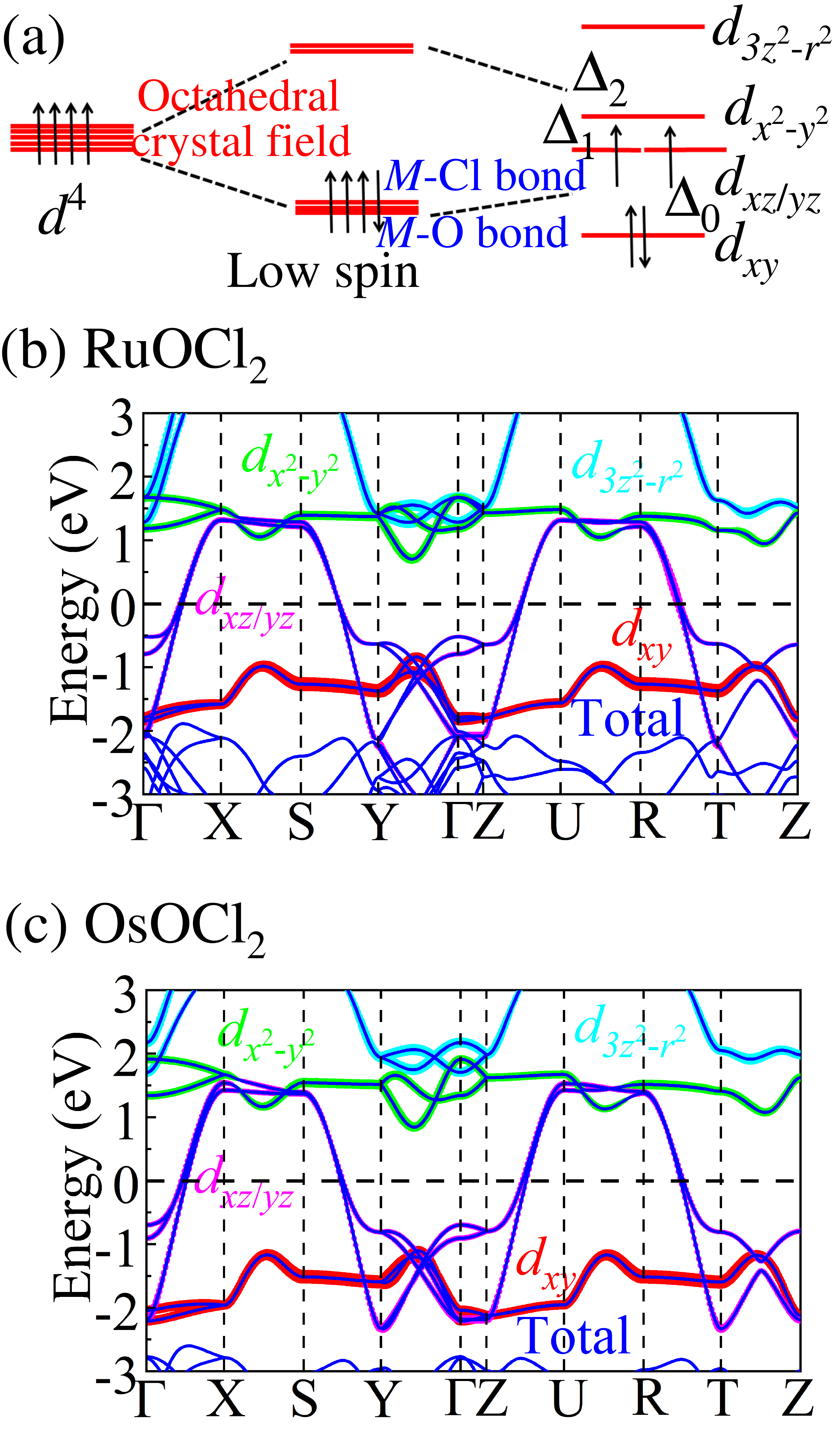}
\caption{(a) Energy splitting of the total $S=1$ $d^4$ electronic configuration. (b-c) Projected band structures of the undistorted I${\rm mmm}$ non-magnetic
phase for (b) RuOCl$_2$ and (c) OsOCl$_2$, respectively. Note that the local \{$x$, $y$, $z$\} axes of projected orbitals are marked in
Fig.~\ref{Fig1}, where the $z$-axis is the $a$-axis and $x$ or $y$ axes are along the $M$-Cl directions. The weight of each Ru or Os orbital is represented by the size of the (barely visible) circles.}
\label{Fig3}
\end{figure}

How do we understand why the FE and Peierls distortions disappeared in RuOCl$_2$ and OsOCl$_2$?
This can be easily understood in the following way: In the octahedral crystal-field, the five $d$ orbitals split into two higher-energy $e_g$ ($d_{x^2-y^2}$ and $d_{3z^2-r^2}$) and three lower-energy $t_{2g}$ ($d_{xy}$, $d_{yz}$, and $d_{xz}$) orbitals, as in Fig.~\ref{Fig3}~(a). Due to their reduced coupling $J_H$, the $4d^4$ and $5d^4$ atoms always favor the total $S=1$ electronic configuration when the $t_{2g}$ orbitals are more than half-filled. Then, replacing the Cl atoms of the octahedral apex by O atoms, it would induce two shortened $M$-O bonds ($M$ = Ru or Os) along the $a$-axis ($z$-axis) and four enlongated $M$-Cl bonds ($M$ = Ru or Os along the $bc$ ($xy$) plane in the $M$O$_2$Cl$_4$ ($M$ = Ru or Os) octahedral configuration, leading to the splitting between the $t_{2g}$ orbitals [see Fig.~\ref{Fig3} (a)]. In this case, the four electrons form a total $S = 1$ electronic configuration with one fully-occupied lower $d_{xy}$ orbital
and two half-filled higher energy $d_{xz/yz}$ bands.

Because two electrons occupy two $d_{xz/yz}$ orbitals in $M$OCl$_2$ ($M$ = Ru or Os), the FE distortion along the $a$-axis is energetically unfavorable, as discussed in VOI$_2$ with the $d^1$ configuration~\cite{Zhang:prb21}. Furthermore, the spin-singlet formation using $d_{xy}$ orbitals
would also be suppressed along the $b$-axis since $d_{xy}$ is a double-occupied state in $M$OCl$_2$ ($M$ = Ru or Os). Hence, both FE and dimerized instabilities are suppressed for RuOCl$_2$ and OsOCl$_2$, resulting in a stable undistorted I${\rm mmm}$ phase (No. 71).

\subsection{B. Electronic properties of the NM state}

Next, we focus on the electronic structures of bulk $M$OCl$_2$ ($M$ = Ru or Os) for the NM state without SOC. Figures~\ref{Fig3}(b) and (c) show that the $e_g$ ($d_{x^2-y^2}$ and $d_{3z^2-r^2}$) bands of Ru's $4d$ and Os's $5d$ orbitals are located at high energy, and therefore are unoccupied. The local $z$-axis is the $a$-axis, while the local $x$ or $y$ axis is along the $M$-Cl directions [see Fig.~\ref{Fig1}(c)], leading to a $d_{xy}$ orbital lying on the $bc$ plane. In addition, we estimated that the energy splitting $\Delta_{2}$ between $d_{x^2-y^2}$ and $d_{3z^2-r^2}$ orbitals is about $1.7$ and $1.6$ eV for RuOCl$_2$ and OsOCl$_2$, respectively, by the weight-center positions of the energy bands. Furthermore, two $4d/5d$ electrons occupy the $d_{xy}$ bands that show only weak dispersion and are far away from the Fermi level. The other two $4d/5d$ electrons of Ru or Os occupy the $d_{xz}$ and $d_{yz}$ orbitals, contributing to the Fermi surface. We also estimated that the energy splitting $\Delta_{1}$ (between $d_{x^2-y^2}$ and $d_{xz/yz}$) and $\Delta_{0}$ (between $d_{xy}$ and $d_{xz/yz}$) are $\Delta_{1}$ = $1.2/1.3$ eV and $\Delta_{0}$ = $1.2/1.4$ eV for RuOCl$_2$ and OsOCl$_2$, respectively. Considering the large crystal-field splitting $\Delta_{0}$, the $J = 0$ singlet ground state induced by SOC may be suppressed in $M$OCl$_2$ ($M$ = Ru or Os), as discussed for the OsCl$_4$ case with the $d^4$ electronic configurations~\cite{Zhang:apl}.

Due to the fully-occupied $d_{xy}$ state and large energy splitting $\Delta$, the magnetic properties of $M$OCl$_2$ ($M$ = Ru or Os) are dictated by the $d_{xz}$ and $d_{yz}$ orbitals to be discussed in the following sections. Moreover, near the Fermi level, $M$OCl$_2$ ($M$ = Ru or Os) displays strongly quasi-1D electronic behavior with contributions from the $d_{xz}$ and $d_{yz}$ orbitals, and the band
structures are much more dispersive along the $a$-axis (i.e. $\Gamma$-X path) than along other directions (i.e. X-S and $\Gamma$-Z paths). For this reason, these materials can be regarded
as ``effective 1D'' systems, as mentioned before.

\begin{figure}
\centering
\includegraphics[width=0.48\textwidth]{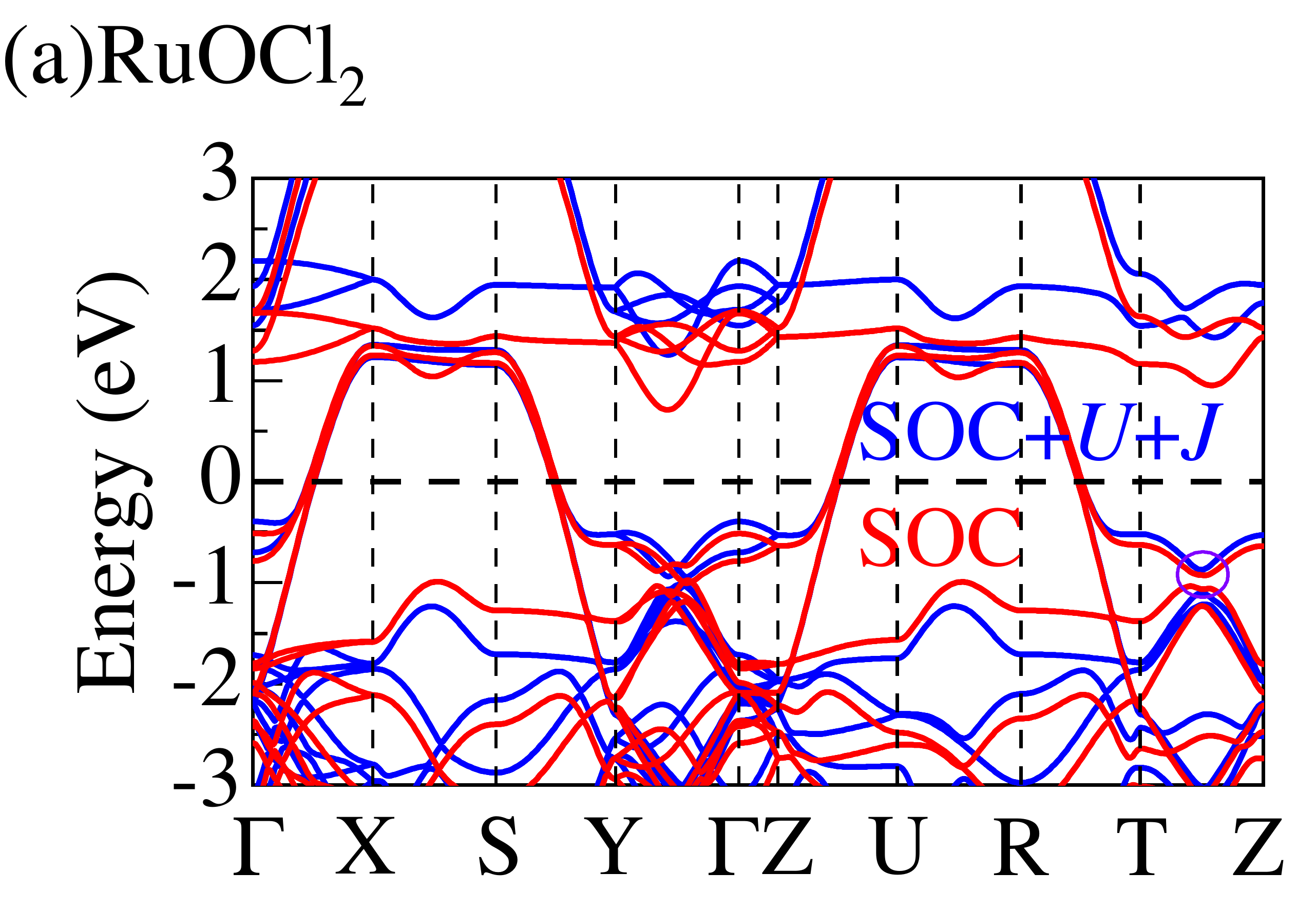}
\includegraphics[width=0.48\textwidth]{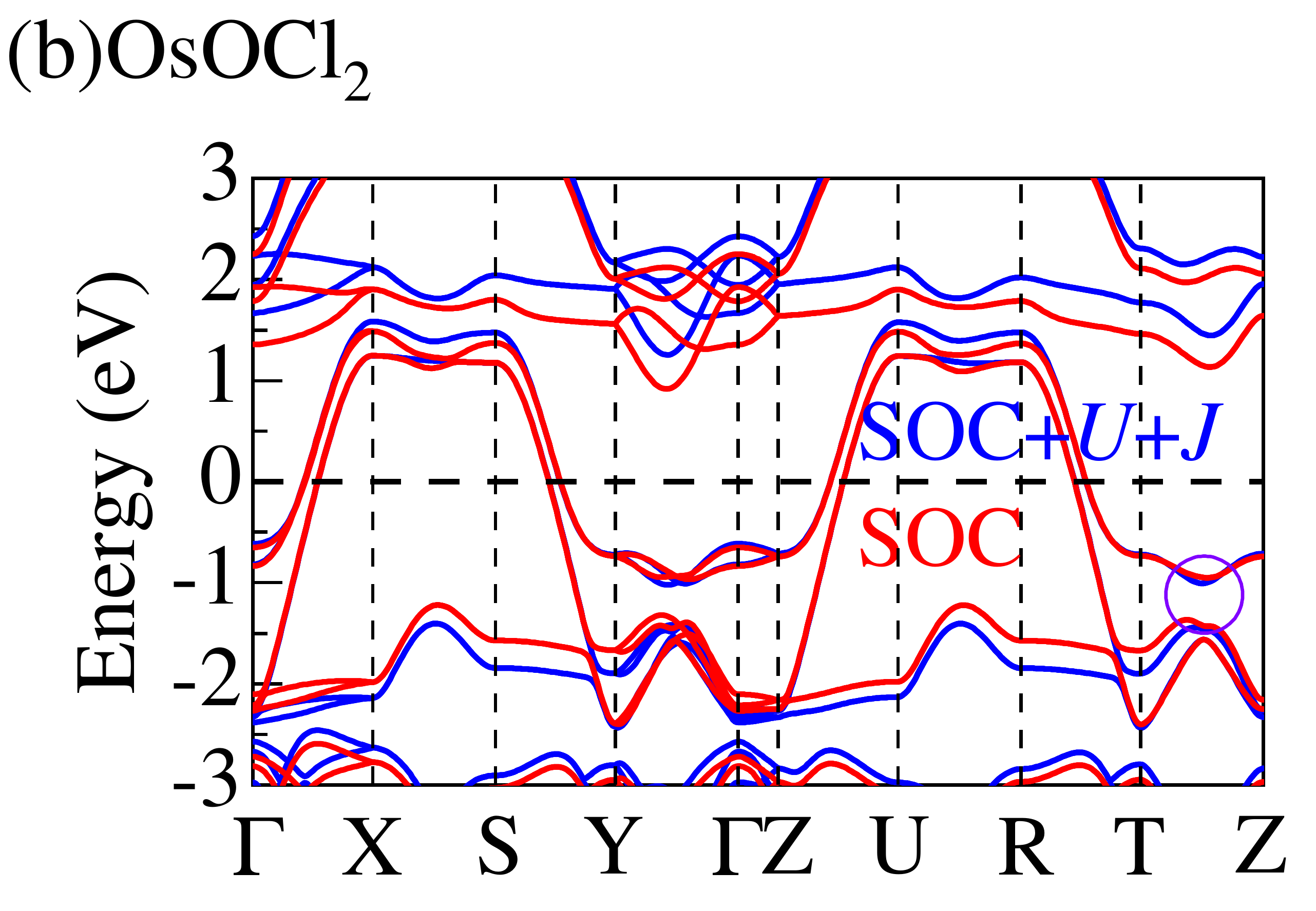}
\caption{(a) Band structure of RuOCl$_2$ in the NM state with SOC and with SOC+$U$+$J$ ($U$ = 3 eV, $J$ = 0.6 eV). (b) Band structure of OsOCl$_2$ in the NM state with SOC and with SOC+$U$+$J$ ($U$ = 2 eV, $J$ = 0.4 eV). The Fermi level is the horizontal dashed line. The energy gaps at about $-1$ eV below the Fermi level are indicated with circles.}
\label{Fig4}
\end{figure}

After considering SOC in the NM state of $M$OCl$_2$ ($M$ = Ru or Os), the bands begin to split as displayed in
Fig.~\ref{Fig4}, opening an energy gap ($\sim 0.12$ eV for RuOCl$_2$ and $\sim 0.41$ eV for OsOCl$_2$) along the T-Z high-symmetry path
at $\sim -1$ eV below the Fermi level. However, introducing SOC, the band structures near the Fermi level do not change much.
Furthermore, the large bandwidth of the $d_{xz/yz}$ states of $M$OCl$_2$ ($M$ = Ru or Os) suggests a large value for the NN hopping of Os or Ru atoms along the $a$-axis. Considering the typical SOC value of Ru and Os atoms~\cite{Zhang:prb22,SOC}, the $J_{\rm eff}$ physics would be suppressed by the large crystal-field splitting and large hopping $t$. In this case, this system should be a spin-1 system, instead of a $J = 0$ singlet ground state, leading to the quenched orbital magnetic moment in those $d^4$ systems, as discussed in the following section. By considering the typically reduced Hubbard $U$ repulsion in 4$d$ and 5$d$ atoms, as compared to 3$d$ atoms, and the large bandwidth of these system (with the hopping $t$ providing the scale), this system is an ``intermediate'' electronic correlation system. Furthermore, we also considered the electronic correlations on Ru ($U = 3$ eV and $J = 0.6$ eV) or Os ($U = 2$ eV and $J = 0.4$ eV) sites, by using the LSDA+$U$ method with Liechtenstein format within the double-counting term~\cite{Liechtenstein:prb}. Figure~\ref{Fig4} also indicates that the lower-energy $d_{xy}$ states of $M$OCl$_2$ ($M$ = Ru or Os) begin to shift away from the Fermi level with fully-occupied characteristics when electronic correlations on the Ru or Os sites are considered.

\begin{figure}
\centering
\includegraphics[width=0.48\textwidth]{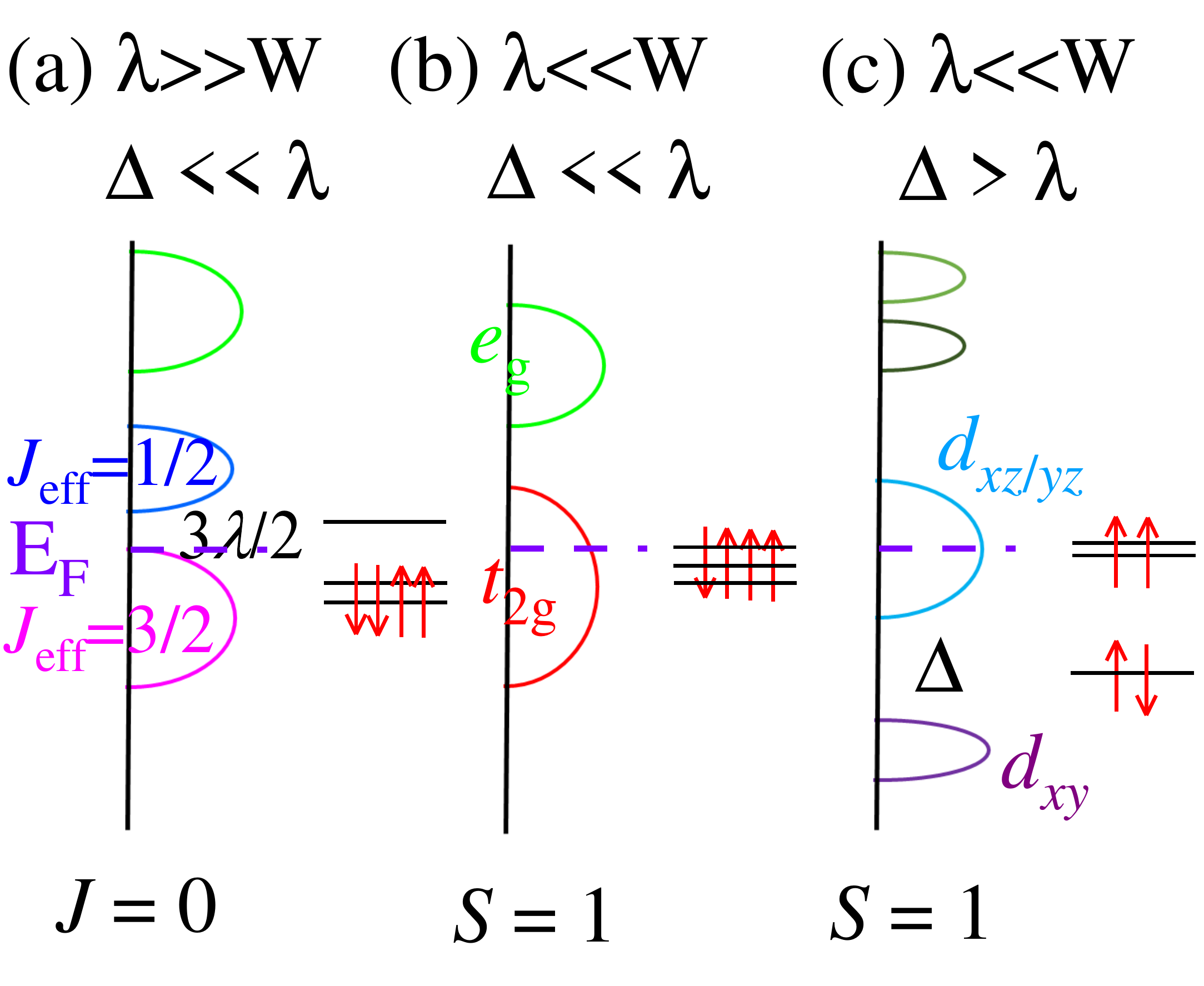}
\caption{The qualitative evolution of the $d^4$ electronic states of Ru$^{4+}$ or Os$^{4+}$. Here, we consider the SOC strength $\lambda$, the bandwidth $W$ (corresponding to $t$), and crystal field splitting $\Delta$ between $d_{xy}$ and $d_{xz/yz}$ orbitals (corresponding to the Jahn-Teller distortion $Q_3 \textless 0$ in an octahedra). (a) $\lambda$ $\gg W$ and $\Delta$  $\ll$ $\lambda$, where a $J = 0$ singlet state is realized. (b)$\lambda$ $\ll W$ and $\Delta$  $\ll$ $\lambda$, where a $S = 1$ state with four electrons in three degenerate $t_{2g}$ orbitals is obtained. (c) $\lambda$ $\ll W$ and $\Delta$  $\textgreater$ $\lambda$, where a $S = 1$ state with one fully-occupied ($d_{xy}$) and two half-occupied $d_{xz/yz}$ orbitals is realized.}
\label{Fig5}
\end{figure}

For the benefit of the readers, we construct a qualitative physical picture for the breakdown of the $J = 0$ singlet ground state in this system, as shown in Fig.~\ref{Fig5}. First, let us discuss the three nearly degenerate $t_{\rm 2g}$ orbitals in a low spin $d^4$ system without electronic correlations. Because $\lambda$ (SOC strength) $\gg W$ (bandwidth, corresponding to the hopping $t$), the system is in a $J = 0$ insulator with fully-occupied $J_{\rm eff}= 3/2$ states [see Fig.~\ref{Fig5}(a)]. In this case, the gap is opened by the splitting between $J_{\rm eff}= 3/2$ and $J_{\rm eff}= 1/2$ states caused by SOC, as displayed in Fig.~\ref{Fig5}(a). If, however, $\lambda$ $\ll W$, then, the system will keep a $S = 1$ state due to the Pauli rule, as shown in Fig.~\ref{Fig5}(b). In this state, four electrons occupy three degenerate $t_{\rm 2g}$ orbitals, leading to a metallic phase. In our case (RuOCl$_2$ and OsOCl$_2$), it is also a $S = 1$ state with one fully-occupied ($d_{xy}$) and two half-occupied ($d_{xz}$ and $d_{yz}$) orbitals, as presented in Fig.~\ref{Fig5}(c). Then, the $J = 0$ singlet ground state is suppressed in our case by the large crystal-field splitting ($\Delta_0$) and large bandwidth ($W$). Finally, as we will discuss in the following section, the system will be a Mott insulator when electron correlations are considered.

\subsection{C. DFT magnetic properties}

\begin{figure}
\centering
\includegraphics[width=0.48\textwidth]{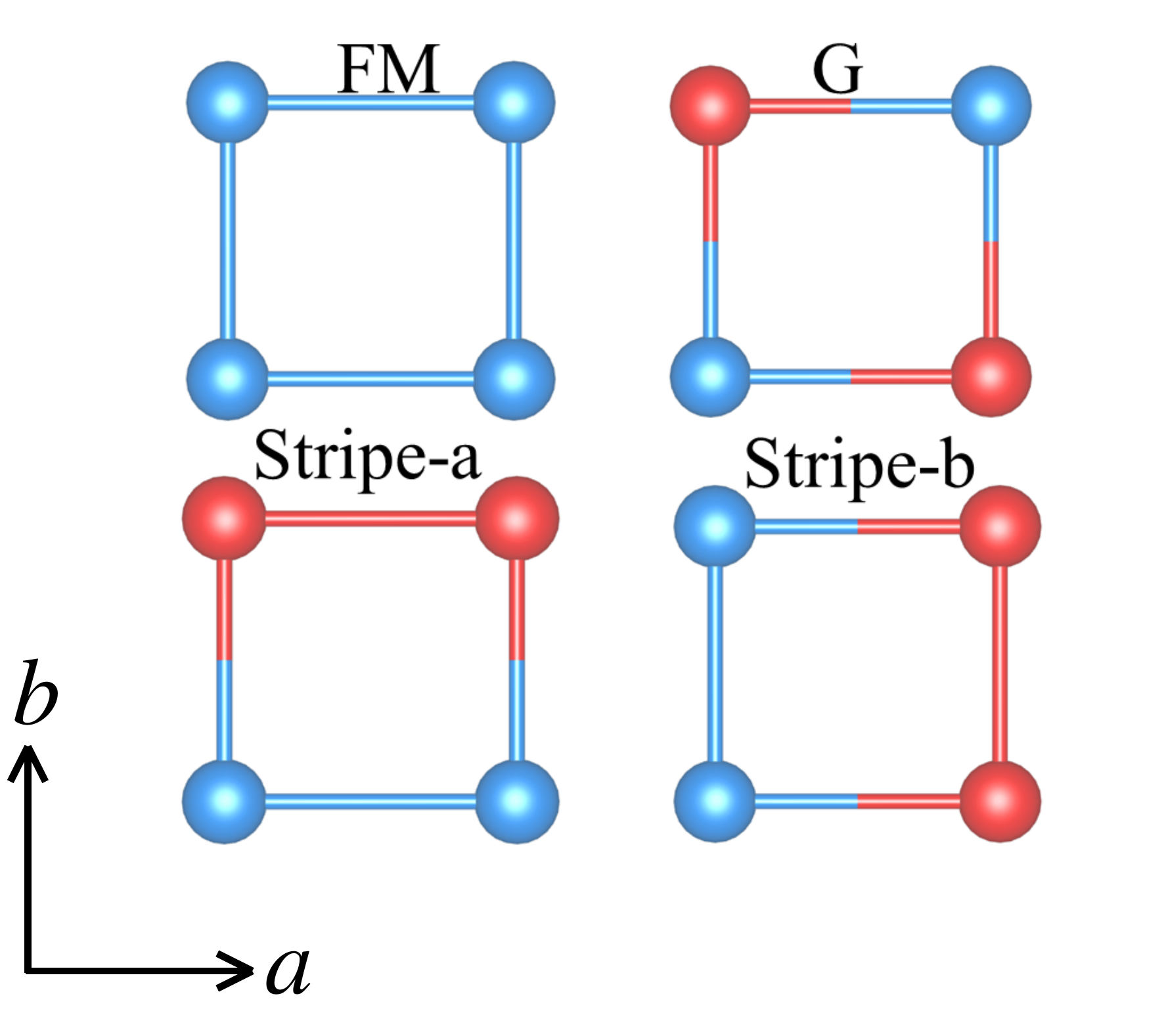}
\caption{Sketch of four possible magnetic patterns in the plane studied here. Spin up and down are indicated by red and blue, respectively.}
\label{Fig6}
\end{figure}

To better understand the in-plane magnetic properties of $M$OCl$_2$ ($M$ = Ru or Os), we also studied several magnetic configurations in a $2\times2\times1$ monolayer structure by considering different NN couplings along the $a$- and $b$- axis, as shown in Fig.~\ref{Fig6}. In addition,  we also relaxed the crystal structures for different spin configurations based on the LSDA+$U$ method with Liechtenstein format~\cite{Liechtenstein:prb}. Here, we used $U = 3$~eV and $J = 0.6$~eV for RuOCl$_2$, and $U = 2$~eV and $J = 0.4$~eV for OsOCl$_2$, respectively, based on previous theoretical studies~\cite{prb11,Vaugier:prb12}.

First, let us discuss the results without SOC, summarized in Table~\ref{Table1}. For RuOCl$_2$, the Stripe-$b$ state has the lowest energy among all tested candidates, while for OsOCl$_2$ the G-type AFM order is the lowest energy. Furthermore, the energy differences between Stripe-b and G-AFM configurations are quite small ($\sim 1 - 2$ meV per Ru or Os), suggesting a weak magnetic exchange coupling along the $M$Cl$_2$ ($M$ = Ru or Os) chain direction, as in our intuitive analysis from the Wannier function results (see Appendix A). In addition, the optimized crystal structures of different spin configurations are very similar to each other, indicating the spin-lattice coupling is not strong in this system. The calculated local spin magnetic moment are about $1.46$ $\mu_{\rm B}$/Ru and $1.02$ $\mu_{\rm B}$/Os for RuOCl$_2$ and OsOCl$_2$, respectively, corresponding to the $S = 1$ configuration in Ru$^{4+}$ or Os$^{4+}$.

\begin{table}
\centering\caption{The optimized lattice constants ({\AA}), local magnetic moments (in $\mu_{\rm B}$/Ru or $\mu_{\rm B}$/Os) within the default PAW sphere, and band gaps (eV) for the various magnetic configurations. Also included are the energy differences (meV/Ru or meV/Os) with respect to the Stripe-b AFM configuration, taken as the reference of energy. All the magnetic states discussed here were fully optimized.}
\begin{tabular*}{0.48\textwidth}{@{\extracolsep{\fill}}lllllc}
\hline
\hline
& Magnetism & $a$/$b$ & $M$ & Gap  & Energy \\
\hline
RuOCl$_2$ &FM     & 3.676/3.564  & 0.002 & 0.19  & 190.3   \\
&G        & 3.709/3.558  & 1.051 & 1.46 & 1.82    \\
&Stripe-a & 3.676/3.562  & 0.177 & 0.28 & 186.5   \\
&Stripe-b & 3.710/3.560  & 1.047 & 1.39 &   0  \\
\hline
OsOCl$_2$ & FM   & 3.727/3.620  & 0.001 & 0.03  & 80.7    \\
&G        & 3.744/3.617  & 0.850 & 1.02 & -0.8    \\
&Stripe-a & 3.727/3.617  & 0.163 & 0.19 & 75.5   \\
&Stripe-b & 3.745/3.621  & 0.845 & 0.89 &   0 \\
\hline
\hline
\end{tabular*}
\label{Table1}
\end{table}

Next, we compared the energies of different spin configurations with SOC. The Stripe-b and G-AFM states still have the lowest energies among all tested candidates for RuOsCl$_2$ and OsOCl$_2$, respectively. Turning on the SOC, the spin quantization axis points to the [010] direction but with only a small difference in energy with respect to the [001] direction, indicating that the spin favors lying in the $bc$ crystal plane, corresponding to the $xy$ plane.  Based on the energy difference between [010] and [001], we obtained that the magnetic anisotropy energies (MAE) are about $1.85$ meV and $18.26$ meV for RuOCl$_2$ and OsOCl$_2$, respectively. Furthermore, the calculated orbital magnetic moment is quenched closed to zero. In this case, the magnetism of this system is almost unaffected by the SOC.

\begin{figure}
\centering
\includegraphics[width=0.48\textwidth]{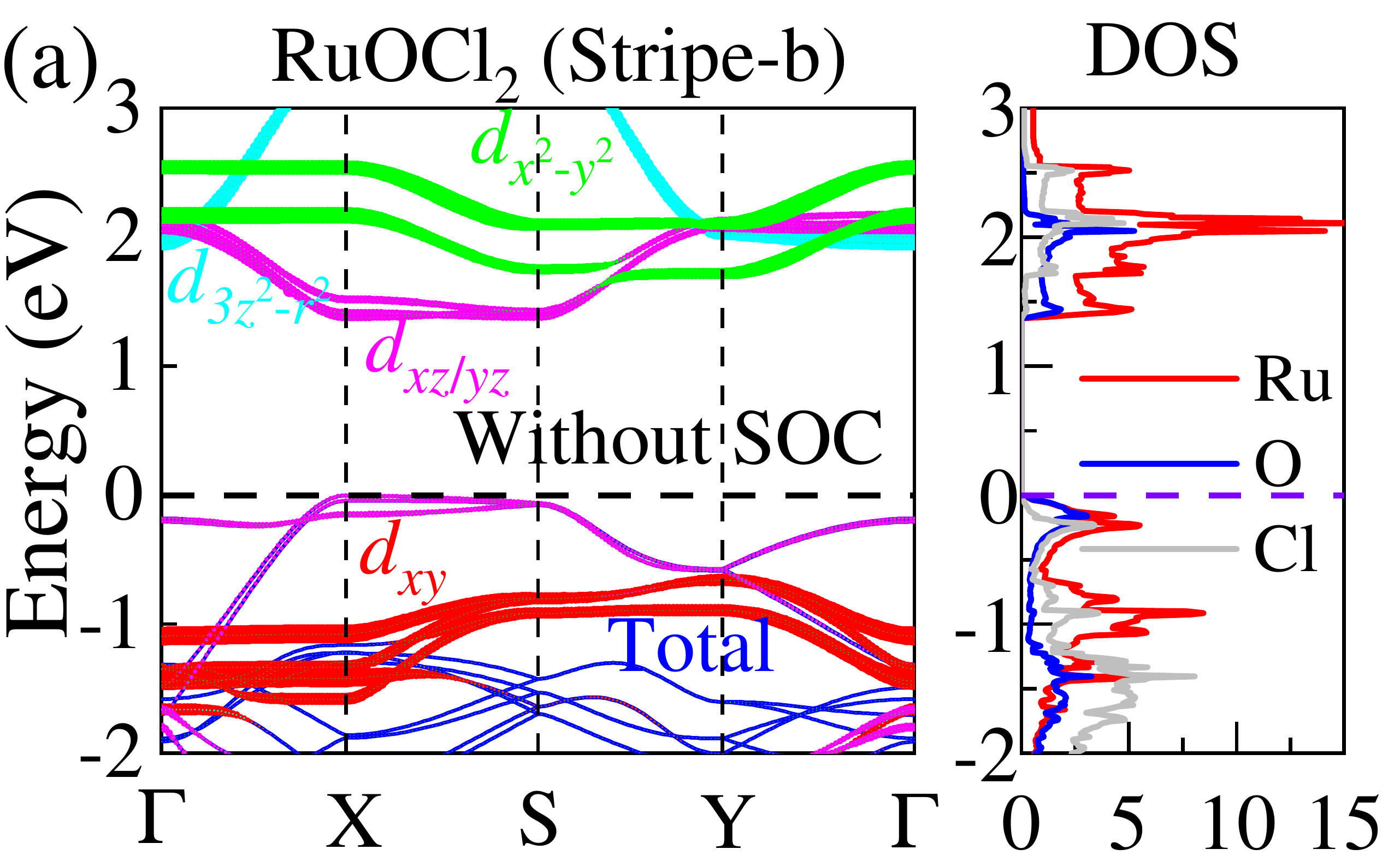}
\includegraphics[width=0.48\textwidth]{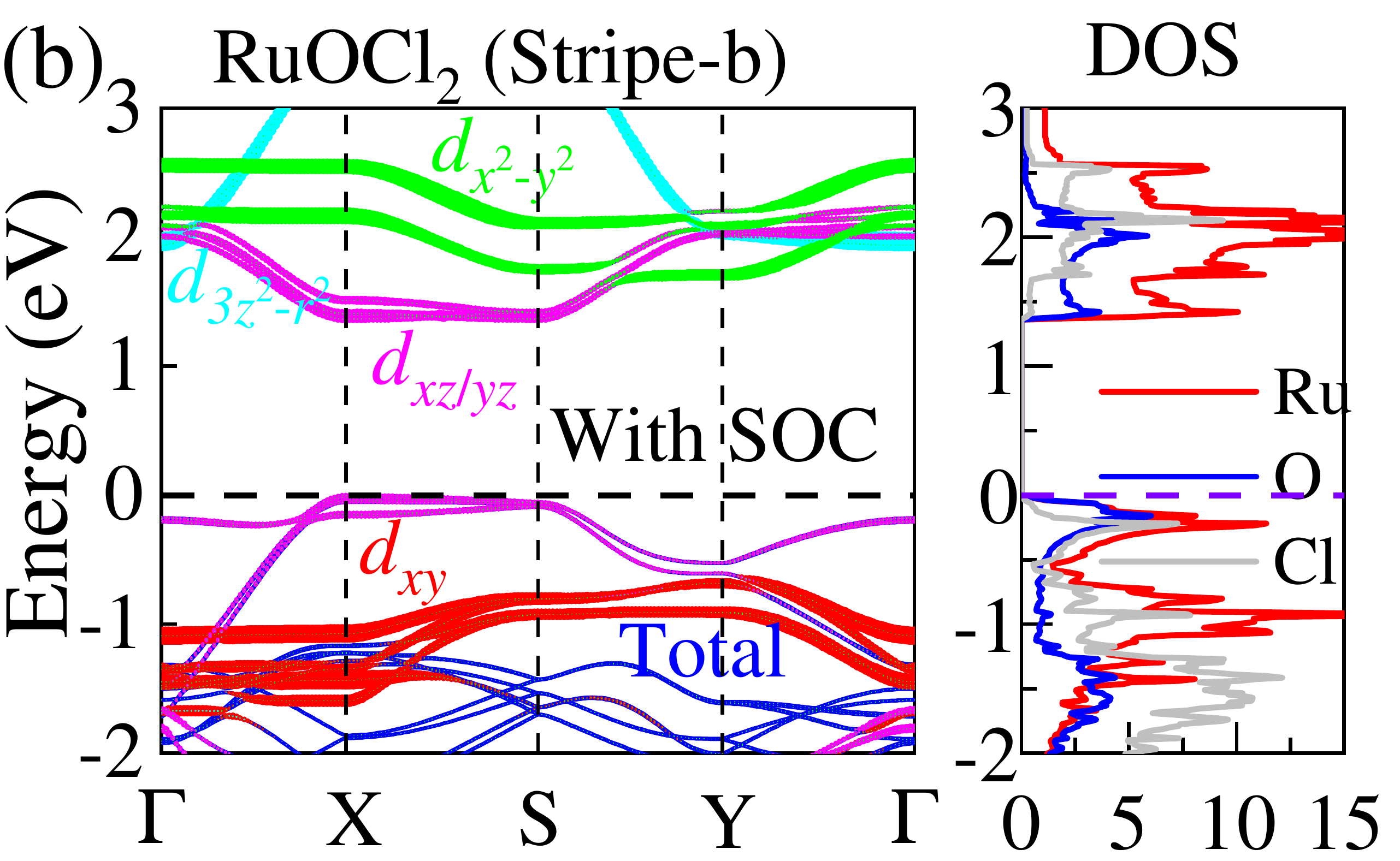}
\caption{Projected band structures and density of states for the Stripe-b state of RuOCl$_2$ (a) without SOC and (b) with SOC, respectively. The Fermi level is shown with dashed horizontal lines. The coordinates of the high-symmetry points in the BZ are $\Gamma$ = (0, 0, 0), X = (0.5, 0, 0), S = (0.5, 0.5, 0), and Y = (0, 0.5, 0).}
\label{Fig7}
\end{figure}

\begin{figure}
\centering
\includegraphics[width=0.48\textwidth]{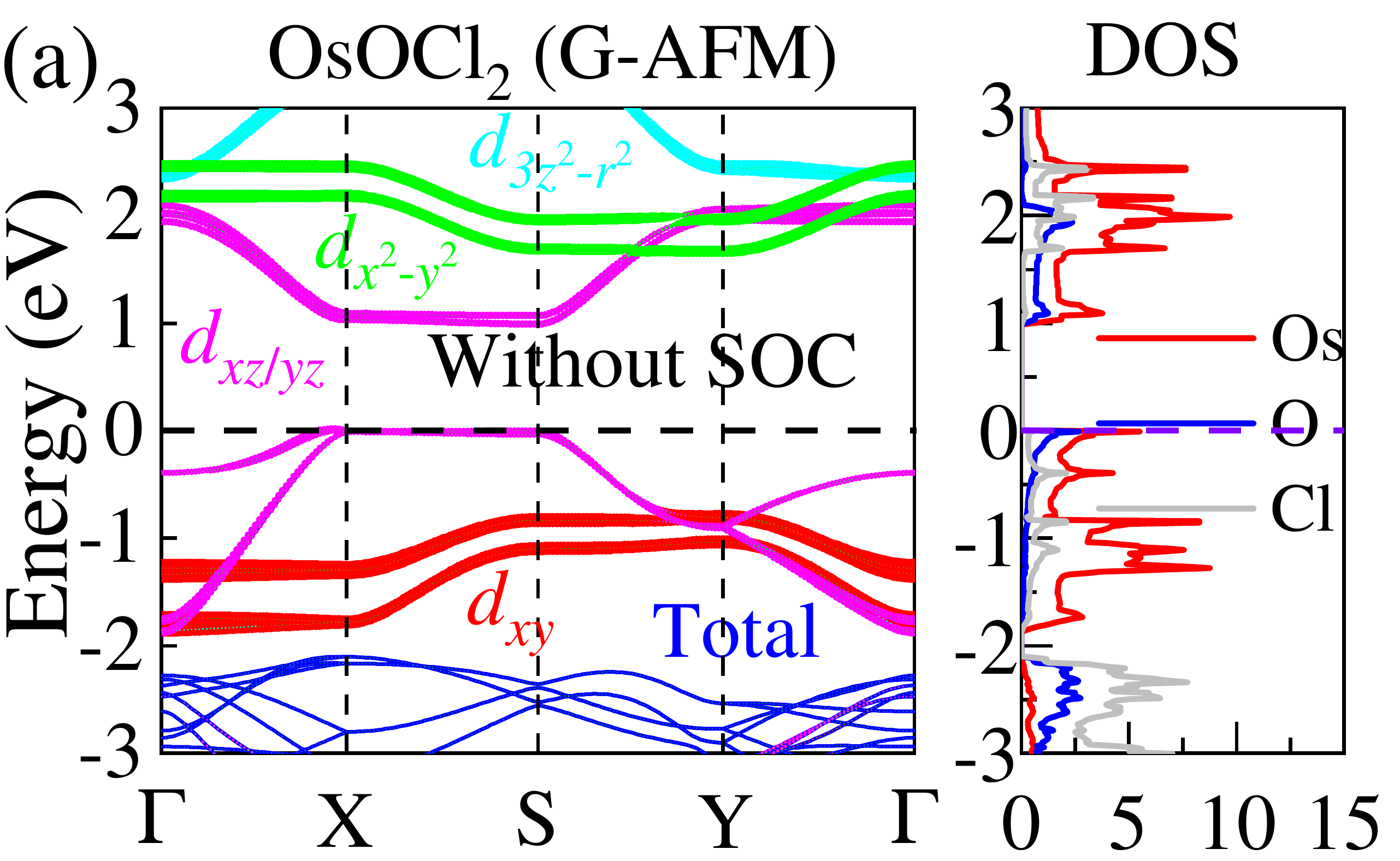}
\includegraphics[width=0.48\textwidth]{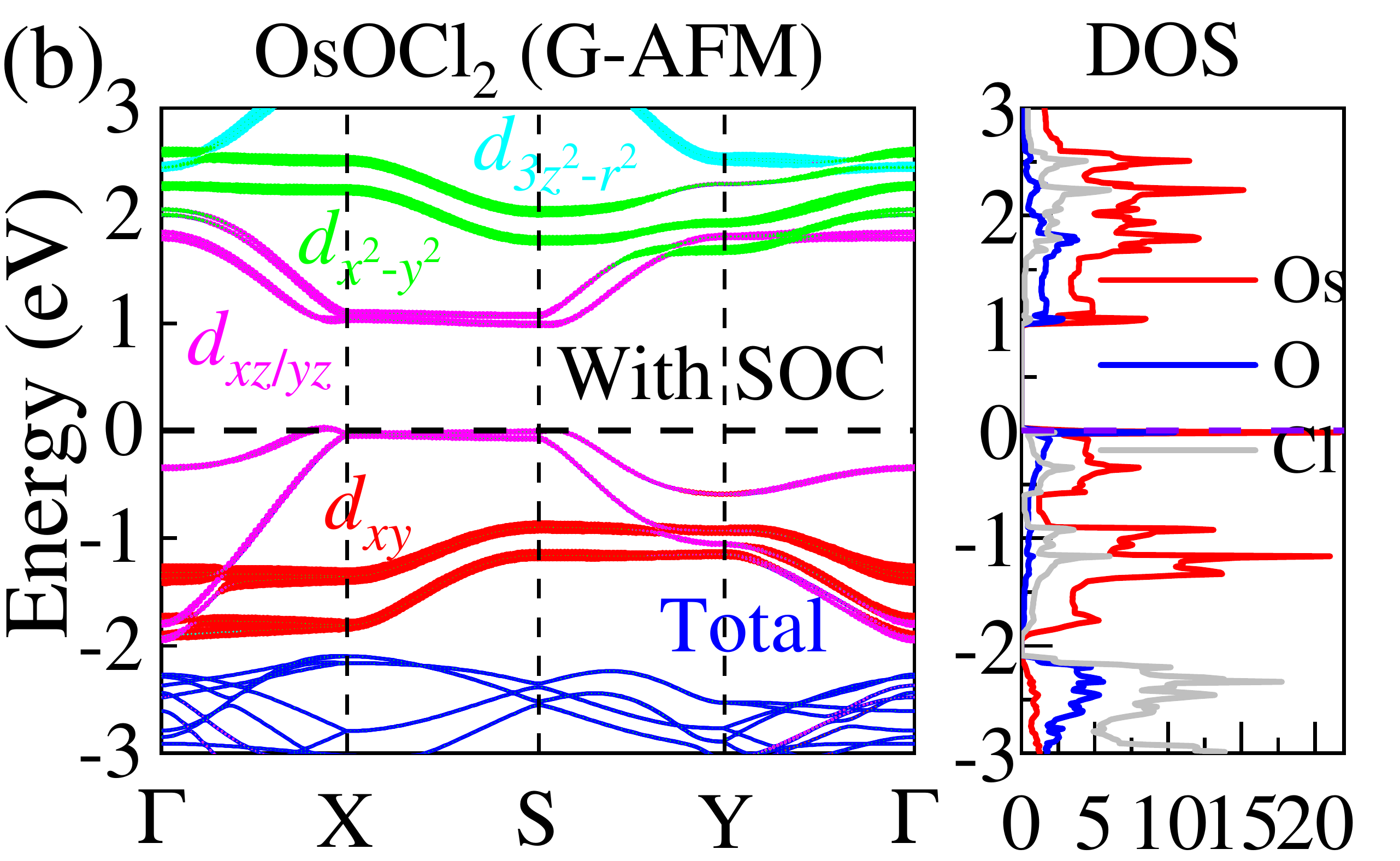}
\caption{Projected band structures and density of states for the G state of OsOCl$_2$ (a) without SOC and (b) with SOC, respectively. The Fermi level is shown with dashed horizontal lines. The coordinates of the high-symmetry points in the BZ are $\Gamma$ = (0, 0, 0), X = (0.5, 0, 0), S = (0.5, 0.5, 0), and Y = (0, 0.5, 0).}
\label{Fig8}
\end{figure}

In Fig.~\ref{Fig7}, we show the band structures of the Stripe-b AFM phase of RuOCl$_2$ calculated with or without SOC.
Figure~\ref{Fig7}(a) indicates that the half-occupied $d_{xz/yz}$ orbitals display Mott-insulating behavior with a gap $\sim 1.4$ eV,
while the $d_{xy}$ orbital is fully-occupied. In this case, this system is in a total $S= 1$ state, where the magnetism
is contributed by the $d_{xz/yz}$ states. Turning on the SOC, the bands begin to split at some high-symmetry points. In addition,
we also calculated the band structure of the G-AFM state of OsOCl$_2$ without or with SOC, as displayed in Fig.~\ref{Fig8}.
Similar to RuOCl$_2$, the $d_{xz/yz}$ orbitals show strong Mott-insulating behavior with a smaller gap $\sim$1 eV, while the fully-occupied
$d_{xy}$ orbital does not contribute to the magnetism. However, the band splitting under SOC is stronger in OsOCl$_2$ than in RuOCl$_2$, as shown in Fig.~\ref{Fig8}(b), considering the Os atom column in the periodic table. Furthermore, the band structure of the magnetic systems is strongly anisotropic along the $a$-axis due to the strongly anisotropic $d_{xz/yz}$ orbitals, leading to an``effective 1D'' magnetic system.

In addition, we also studied the effect of different values of $U$ (from 2 to 4 eV) without SOC, where $J$ is chosen as 0.6 or 0.4 eV for RuOCl$_2$ and OsOCl$_2$, respectively.
The optimized crystal lattices are almost unchanged: for example, $a = 3.772$ and $b = 3.556$ \AA~for RuOCl$_2$ at $U = 4$~eV.
By considering different effective $U$'s, the Stripe-b and G-AFM states still have the lowest energies among all the tested candidates, with only a tiny difference in energy between RuOCl$_2$ and OsOCl$_2$.
For RuOCl$_2$, the Stripe-b AFM state has the lowest energy from $U = 2$~eV to 4~eV and the energy difference between Stripe-b and G-AFM increases from 1.13 meV at $U = 2$~eV to 2.29 meV at $U = 4$~eV.
For RuOCl$_2$, the G-type AFM has the lowest energy at $U = 2$ and 3~eV, while the Stripe-b AFM order has the lowest energy at $U = 4$~eV.

\begin{table}
\centering\caption{The calculated total energy (in meV) per Ru or Os with different spin orientations (along [100], [010] and [001] crystal axes) and magnetic moments (in $\mu_{\rm B}$/Ru or $\mu_{\rm B}$/Os =  units) for the ground state of RuOCl$_2$ and OsOCl$_2$. The total energy with [100] spin orientation ($a$ crystal axis) is set to zero. MAE (in mev) per Ru or Os is obtained by comparing the energy difference between [010] and [001]. Here, we used different $U$ (from 2 to 4 eV) at $J = 0.6$ eV and $J = 0.4$ eV for  RuOCl$_2$ and OsOCl$_2$, respectively.}
\begin{tabular*}{0.48\textwidth}{@{\extracolsep{\fill}}llllllc}
\hline
\hline
&E(100) & E(010) & E(001)& m(spin)  & m (orbial) & MAE\\
\hline
RuOCl$_2$\\
\hline
U = 2 eV & 0 & -1.96 & -1.96  & 0.878 & 0.003 & 1.96    \\
U = 3 eV & 0 & -1.85 & -1.85  & 1.052 & 0.002 & 1.85    \\
U = 4 eV & 0 & -1.71 & -1.70  & 1.178 & 0.002 & 1.71    \\
\hline
OsOCl$_2$\\
\hline
U = 2 eV &0 &  -18.26 & -17.89& 0.842 & 0.004  & 18.26    \\
U = 3 eV &0 &  -17.41 & -16.95& 1.081 & 0.004  & 16.95    \\
U = 4 eV &0 &  -15.81 & -15.74& 1.262 & 0.008  & 15.74    \\
\hline
\hline
\end{tabular*}
\label{Table2}
\end{table}

\begin{figure}
\centering
\includegraphics[width=0.48\textwidth]{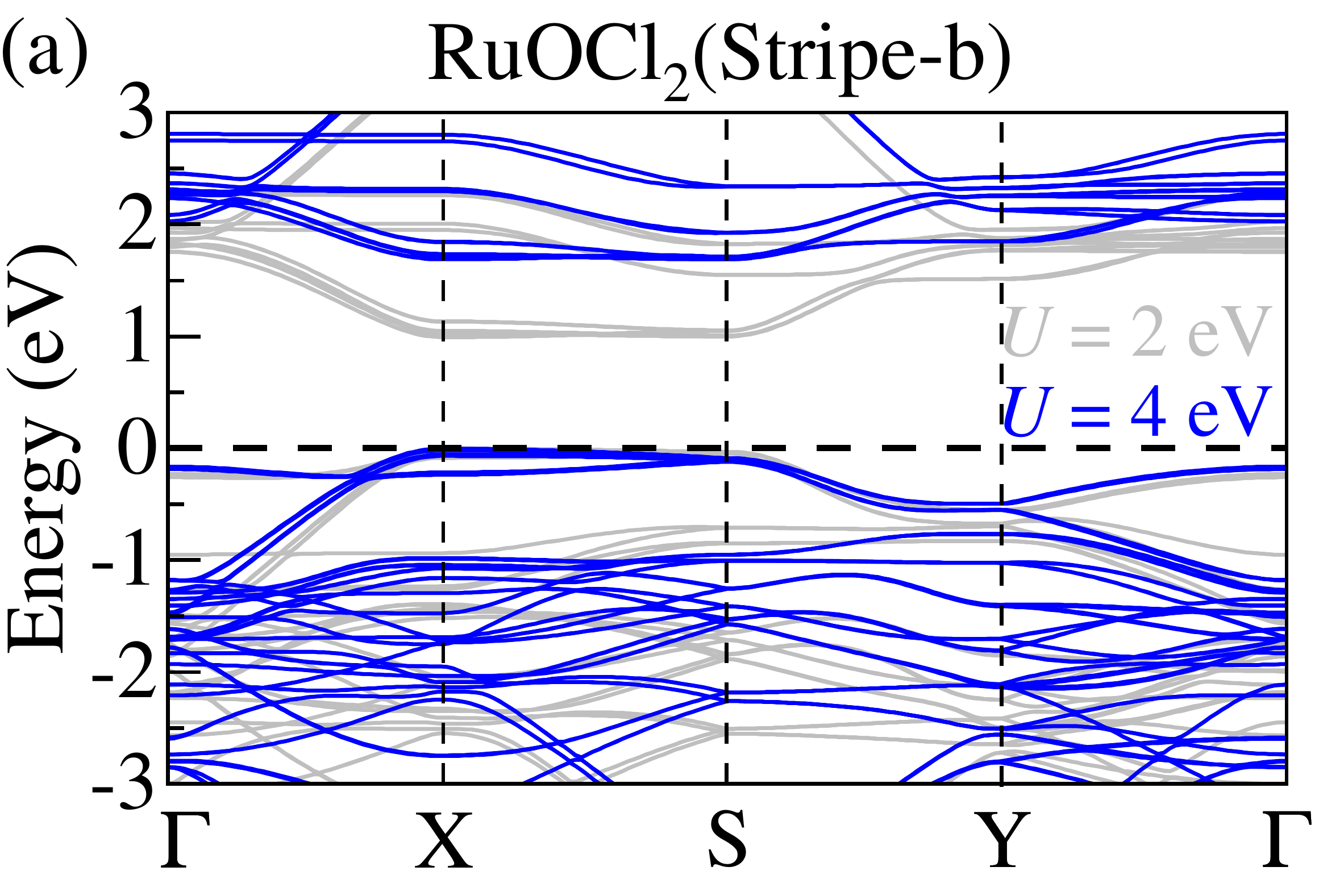}
\includegraphics[width=0.48\textwidth]{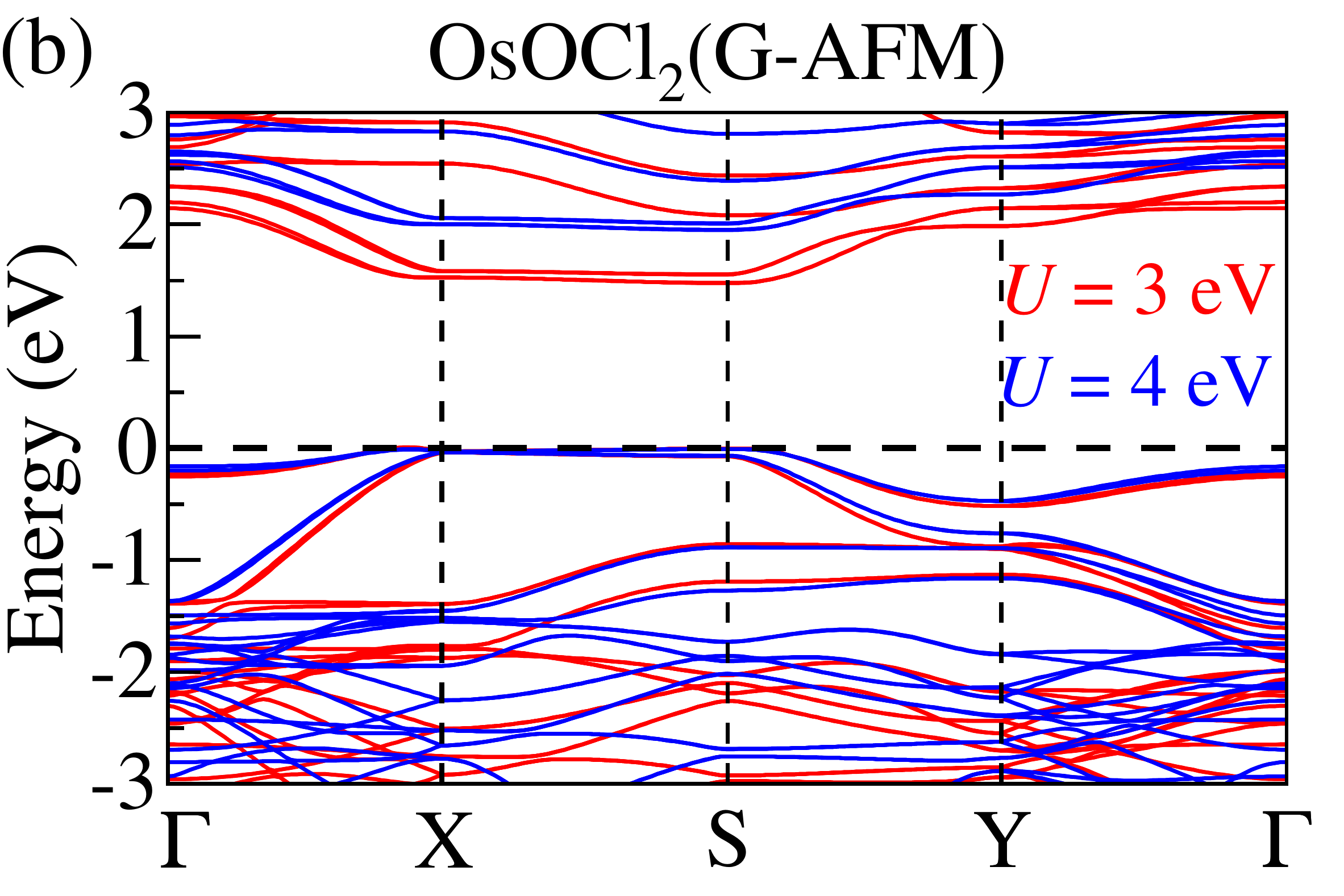}
\caption{Band structures for (a) the Stripe-b AFM state of RuOCl$_2$ and (b) the G-AFM state of OsOCl$_2$, both with SOC and $U$. The Fermi level is shown with dashed horizontal lines.}
\label{Fig9}
\end{figure}

Turning on the SOC for other values of $U$, the spin quantization axis still points along the [010] direction but with only a small difference in energy with respect to the [001] direction. The spin favors lying in the $bc$ crystal plane, corresponding to the $xy$ plane, independently of the choice of $U$ in the range studied. As summarized in Table~\ref{Table2}, the MAE does not change much for RuOCl$_2$ and OsOCl$_2$ at different $U$'s slightly decreasing as $U$ increases. Furthermore, all the calculated orbital magnetic moments are quenched closed to zero. In addition, we also studied the electronic structures of different $U$'s for RuOCl$_2$ and OsOCl$_2$ with SOC. As shown in Fig.~\ref{Fig9}, the Mott-gap contributed by half-filling the $d_{xz/yz}$ orbitals begins to increases as $U$ increases for both RuOCl$_2$ and OsOCl$_2$, as expected.

Finally, as a side remark, note that critical temperatures cannot be evaluated with DFT. Moreover, even with the DMRG to be used in the next section, due to the 1D nature of the chains studied, a finite critical temperature can only be obtained after including a weak coupling along the perpendicular directions, a formidable task for DMRG.
Thus, estimations of those critical temperatures are postponed for future work.

\section{IV. Multi-orbital Hubbard Model and DMRG results}

For low-dimensional systems, interesting phenomena caused by strongly anisotropic electronic structures have been qualitatively unveiled in theory by using simple 1D models, including 1D spin order~\cite{Gao:prb20}, orbital ordering~\cite{Pandey:prb21,Lin:prm21}, nodes in the spin density~\cite{Lin:prb21}, as well as dimerization ~\cite{Zhang:prb21,Zhang:ossp}.

In these 1D-effective systems, the quantum fluctuations may be important to clarify the true magnetic ground state properties. Because DFT neglects fluctuations, we constructed an effective multi-orbital Hubbard model and then used DMRG, which includes quantum fluctuations~\cite{white:prl,white:prb,riera97}, to better understand the quasi-1D magnetic behavior of $M$OCl$_2$ ($M$ = Ru or Os) along the dominant $a$-axis in the $d^4$ electronic configuration. The model studied here includes kinetic energy and interaction energy terms $H = H_k + H_{int}$. The tight-binding kinetic portion is described as
\begin{eqnarray}
H_k = \sum_{\substack{i\sigma\gamma\gamma'}}t_{\gamma\gamma'}(c^{\dagger}_{i\sigma\gamma}c^{\phantom\dagger}_{i+1\sigma\gamma'}+H.c.)+ \sum_{i\gamma\sigma} \Delta_{\gamma} n_{i\gamma\sigma}\,,
\end{eqnarray}
where the first part represents the hopping of an electron from orbital $\gamma$ at site $i$ to orbital $\gamma'$ at the NN site $i+1$ on a chain of length $L$. $\gamma$ and $\gamma'$ represent the three different orbitals \{$d_{xz}$, $d_{yz}$, $d_{xy}$\} indexed as $\gamma$ =  \{0, 1, 2\}.

The standard interaction part of the Hamiltonian is given by
\begin{eqnarray}
H_{int}= U\sum_{i\gamma}n_{i\uparrow \gamma} n_{i\downarrow \gamma} +(U'-\frac{J_H}{2})\sum_{\substack{i\\\gamma < \gamma'}} n_{i \gamma} n_{i\gamma'} \nonumber \\
-2J_H  \sum_{\substack{i\\\gamma < \gamma'}} {{\bf S}_{i\gamma}}\cdot{{\bf S}_{i\gamma'}}+J_H  \sum_{\substack{i\\\gamma < \gamma'}} (P^{\dagger}_{i\gamma} P_{i\gamma'}+H.c.)\,.
\end{eqnarray}
The first term describes the intraorbital Hubbard repulsion and the second term the interorbital repulsion, where the standard relation $U'=U-2J_H$ is assumed due to rotational invariance. The third term represents the Hund's coupling between electrons occupying the $d$ orbitals, and the fourth term is the pair-hopping between different orbitals at the same site $i$, where $P_{i\gamma}$=$c_{i \downarrow \gamma} c_{i \uparrow \gamma}$.

As explained above, to solve this multi-orbital Hubbard model and obtain magnetic properties along the $a$-axis
we used DMRG, as implemented in the DMRG++ software~\cite{Alvarez:cpc}. Specifically, we employed a $24$-sites chain with open-boundary conditions (OBC). Furthermore, at least $1400$ states were kept and up to $21$ sweeps were performed during our DMRG calculations. The electronic filling $n = 4$ in the three orbitals was considered. This electronic density (four electrons in three orbitals) corresponds to the total $S=1$ configurations of the $d^4$ configuration of Ru$^{\rm 4+}$ or Os$^{\rm 4+}$.

In the tight-binding term, we only considered the NN hopping matrix of OsOCl$_2$ along the $a$-axis ($M$-O direction). The crystal-field splitting $\Delta$s of orbitals $\gamma$ are also obtained from the Wannier results of OsOCl$_2$. More details about the Wannier functions and hoppings can be found in Appendix B.
The total kinetic energy bandwidth $W$ is 3~eV. To reproduce the data shown in this publication, we prepared notes and input files at~\cite{dmrgplusplus} and supplemental materials~\cite{Supplemental}.

This system can be regarded as an ``effective'' low-energy model with four electrons in three orbitals, corresponding to an electronic density per orbital $4/3$. In addition, the SOC is not important for the magnetism, hence, we do not introduce SOC in our model.
The NN hopping matrix used here is:
\begin{equation}
\begin{split}
t_{\gamma\gamma'} =
\begin{bmatrix}
         -0.713     &   0.013  &   0.000	   	       \\
          0.013     &  -0.717  &   0.000	   	       \\
          0.000	    &   0.000  &  -0.011	
\end{bmatrix}.\\
\end{split}
\end{equation}

\begin{figure}
\centering
\includegraphics[width=0.48\textwidth]{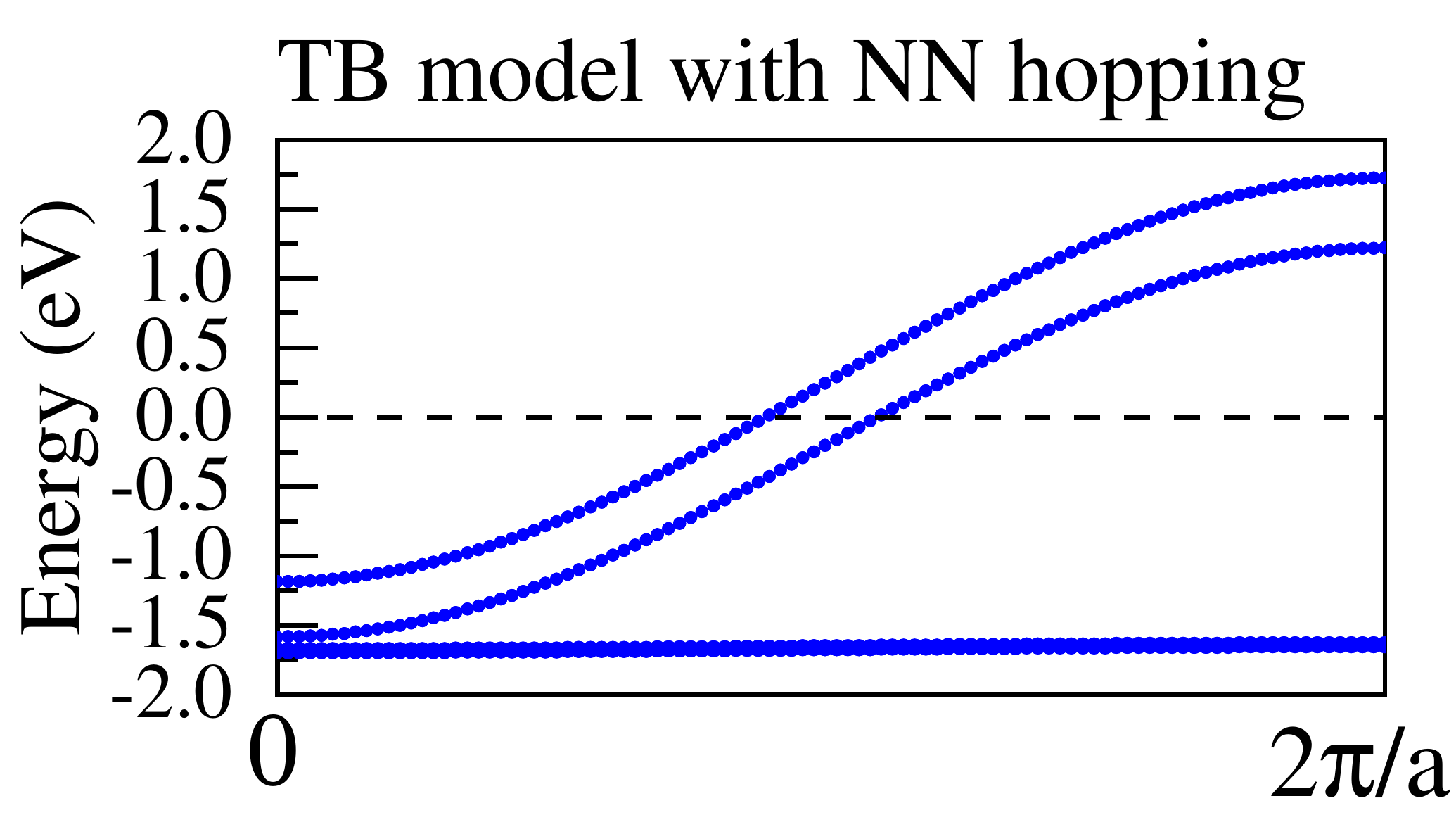}
\caption{Three-orbital tight-binding model with nearest-neighbor hoppings of OsOCl$_2$ along the $a$-axis.
In the text in this section we explain that using three bands is a simplification from the six bands found
in Fig.~\ref{Fig6} near the Fermi level, because in the tight-binding only one Os atom is used, while in the
band structure calculation we employ unit cells with two Os atoms.}
\label{Fig10}
\end{figure}

This reduction in complexity allows us to perform unbiased DMRG calculations for this system. As displayed in Fig.~\ref{Fig10}, the three-orbital tight-binding bandstructure agrees qualitatively with the DFT bandstructure along the $a$-axis. Note that a perfect agreement Wannier and the tight-binding bands for DMRG would require more long-range hoppings.
In our band structure calculation for the nonmagnetic state, see Fig.~\ref{Fig6},
we include {\it two layers} in a unit cell with two Os atoms. Specifically, in Fig.~\ref{Fig6} in the range
from -2.5~eV to 0~eV there are six bands because of the two Os atoms used, each contributing three $t_{2g}$ orbitals.
Meanwhile, in the tight-binding calculation Fig.~\ref{Fig10} there are only three bands because only one Os is used.
Nevertheless, qualitative features related to dominant magnetic states are expected to be captured by this simplification.

Next, we measured several observables by using the three-orbital Hubbard model and the DMRG algorithm. The real-space
spin-spin correlations are defined as
\begin{eqnarray}
S(r)=\langle{{\bf S}_i \cdot {\bf S}_j}\rangle.
\end{eqnarray}
Here $r=\left|{i-j}\right|$, and the spin at site $i$ is
\begin{eqnarray}
{\bf S}_i = \frac{1}{2}\sum_\gamma\sum_{\alpha\beta}c^\dagger_{i\gamma\alpha}{\bf \sigma}_{\alpha\beta}c^{\phantom\dagger}_{i\gamma\beta}\,,
\end{eqnarray}
where ${\bf \sigma}_{\alpha\beta}$ are the matrix elements of the Pauli matrices.
The spin structure factor is defined as
\begin{eqnarray}
S(q)=\frac{1}{L}\sum_{r}e^{-iqr}S(r).
\end{eqnarray}

The site-average occupancy of orbitals is
\begin{eqnarray}
n_{\gamma}=\frac{1}{L}\sum_{\substack{i,\sigma}}{\langle}n_{i\gamma\sigma}\rangle.
\end{eqnarray}

The squared local spin, averaged over all the sites, is
\begin{eqnarray}
\langle S^2 \rangle=\frac{1}{L}\sum_{i}\langle{{\bf S}_i \cdot {\bf S}_i}\rangle.
\end{eqnarray}

\begin{figure}
\centering
\includegraphics[width=0.48\textwidth]{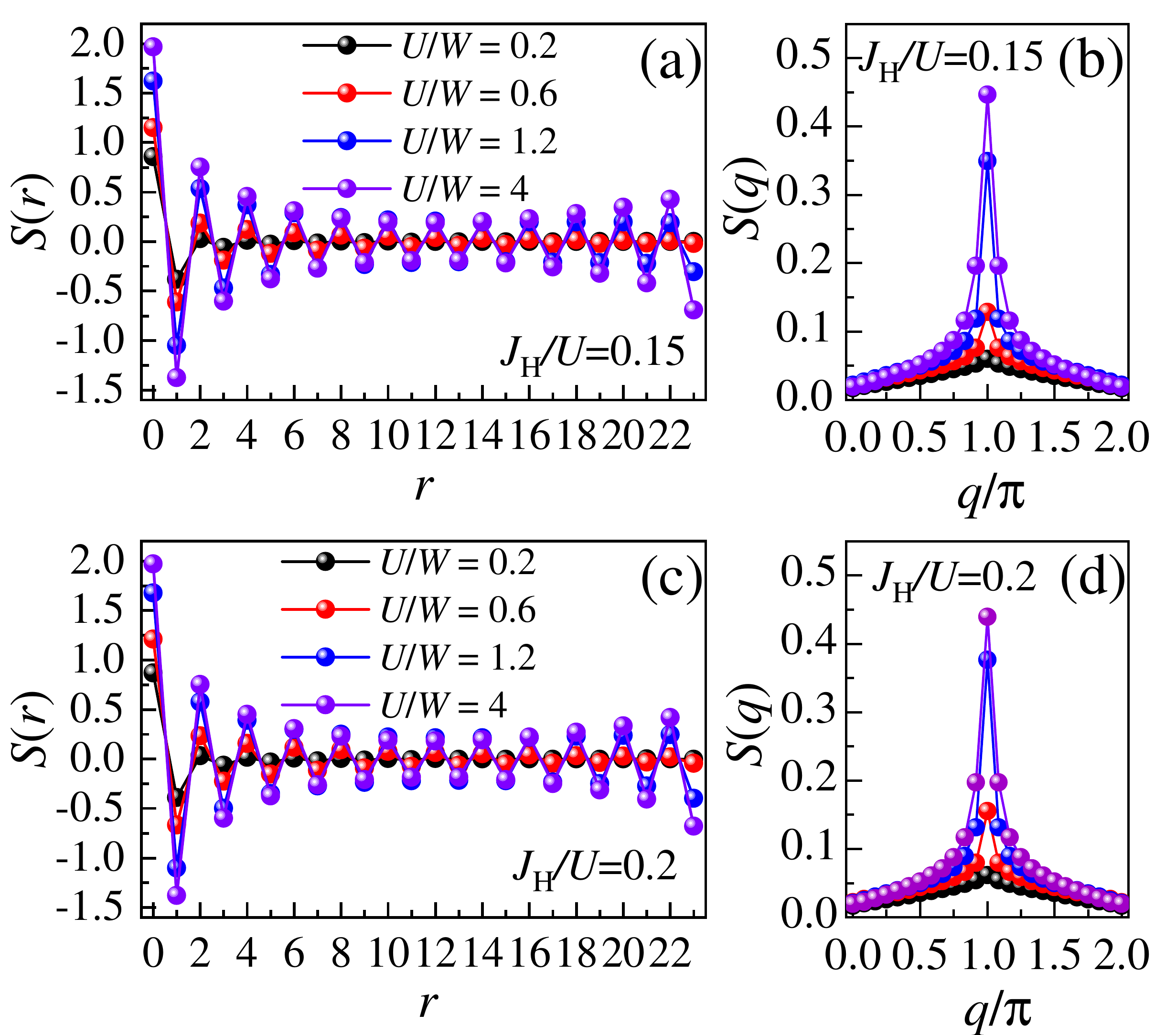}
\caption{Spin-spin correlations $S(r)=\langle{{\bf S}_i \cdot {\bf S}_j}\rangle$ (with $r=\left|{i-j}\right|$ in real space) for $J_H/U=0.15$ (a) and $J_H/U=0.2$ (c). The spin structure factor $S(q)$ for different values of $U/W$ and $J_H$/$U$ = 0.15 (b) and $J_H/U=0.2$ (d). Here we have used a chain of length $L = 24$.}
\label{Fig11}
\end{figure}

Figure~\ref{Fig11} illustrates our calculation of the dominant magnetic coupling along the $a$-axis based on the DMRG measurements of spin-spin correlations and spin structure factors. Panel (a) shows the spin-spin correlation $S(r)$=$\langle{{\bf S}_i \cdot {\bf S}_j}\rangle$ vs. distance $r$, for different values of $U/W$, at $J_H/U = 0.15$. Here, the distance is defined as $r=\left|{i-j}\right|$, with site indices $i$ and $j$. For weak electronic correlations, the system displays paramagnetic (PM) behavior since the spin correlation $S(r)$ decays rapidly with distance $r$ [see the result at $U/W$ = 0.2 in Fig.~\ref{Fig11}(a)]. At $U/W = 0.6$, the spin correlation $S(r)$ indicates weak staggered AFM coupling along the
$M$-O chain direction with a small peak of the spin structure factor $S(q)$ at $q = \pi$, as shown in Fig.~\ref{Fig11}(b). Then, by increasing $U/W$, the spin correlation $S(r)$ shows that the system transfers into the canonical staggered AFM phase with the $\uparrow$-$\downarrow$-$\uparrow$-$\downarrow$ configuration in the whole region of our study ($U/W \le 8$) [see the results at $U/W = 1.2$ and 4 in Fig.~\ref{Fig11}(a)]. As shown in Fig.~\ref{Fig11}(b), the spin structure factor $S(q)$ displays a sharp peak at $q = \pi$ at $U/W = 1.2$ and 4, corresponding to the canonical staggered AFM phase. In addition, we also calculated the spin-spin correlation $S(r)$ and spin structure factor $S(q)$ at $J_H/U = 0.2$, which are similar to the results using $J_H/U = 0.15$, as shown in Figs.~\ref{Fig11}(c) and (d).  Note that in one dimension,
quantum fluctuations prevent full long-range order. But the staggered order tendency is clear at both $J_H/U = 0.15$ and $J_H/U = 0.2$.

In the range of $U/W$ studied, a dominant AFM state was found ($\uparrow$-$\downarrow$-$\uparrow$-$\downarrow$) in our DMRG calculations. This is physically reasonable, considering known facts about the Hubbard model. Based on the hopping matrix from MLWFs calculations, the $\gamma=0$ and $\gamma=1$ orbitals clearly have much larger hopping amplitudes than the $\gamma=2$ orbital, leading to the formation of the AFM order. Furthermore, the diagonal hopping amplitudes are dominant and give rise to the direct exchange mechanism in this system. In this case, the standard superexchange Hubbard spin-spin interaction dominates, causing the spins to order antiferromagnetically along the chain, in agreement with our DFT calculations.

\begin{figure}
\centering
\includegraphics[width=0.48\textwidth]{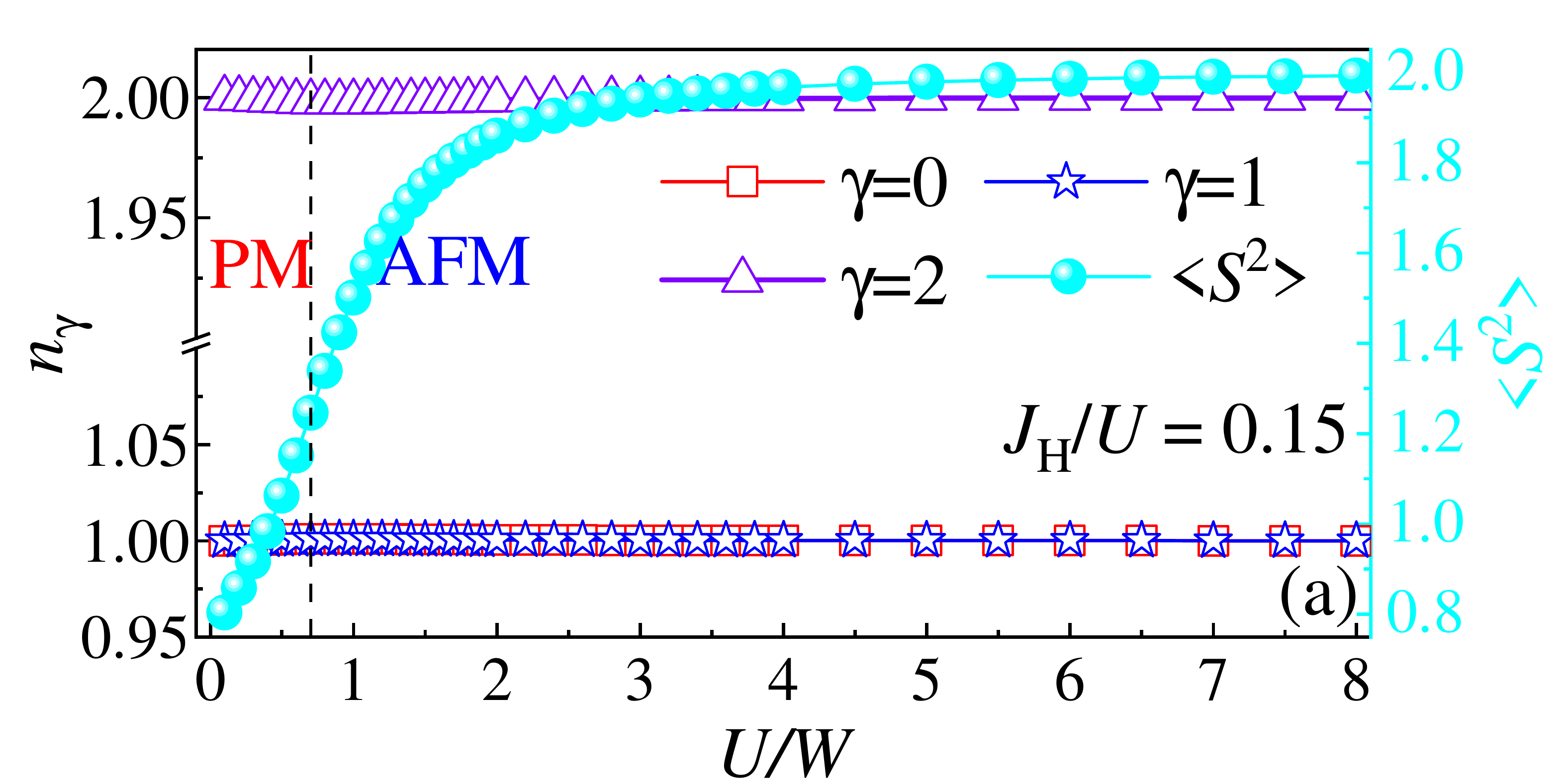}
\includegraphics[width=0.48\textwidth]{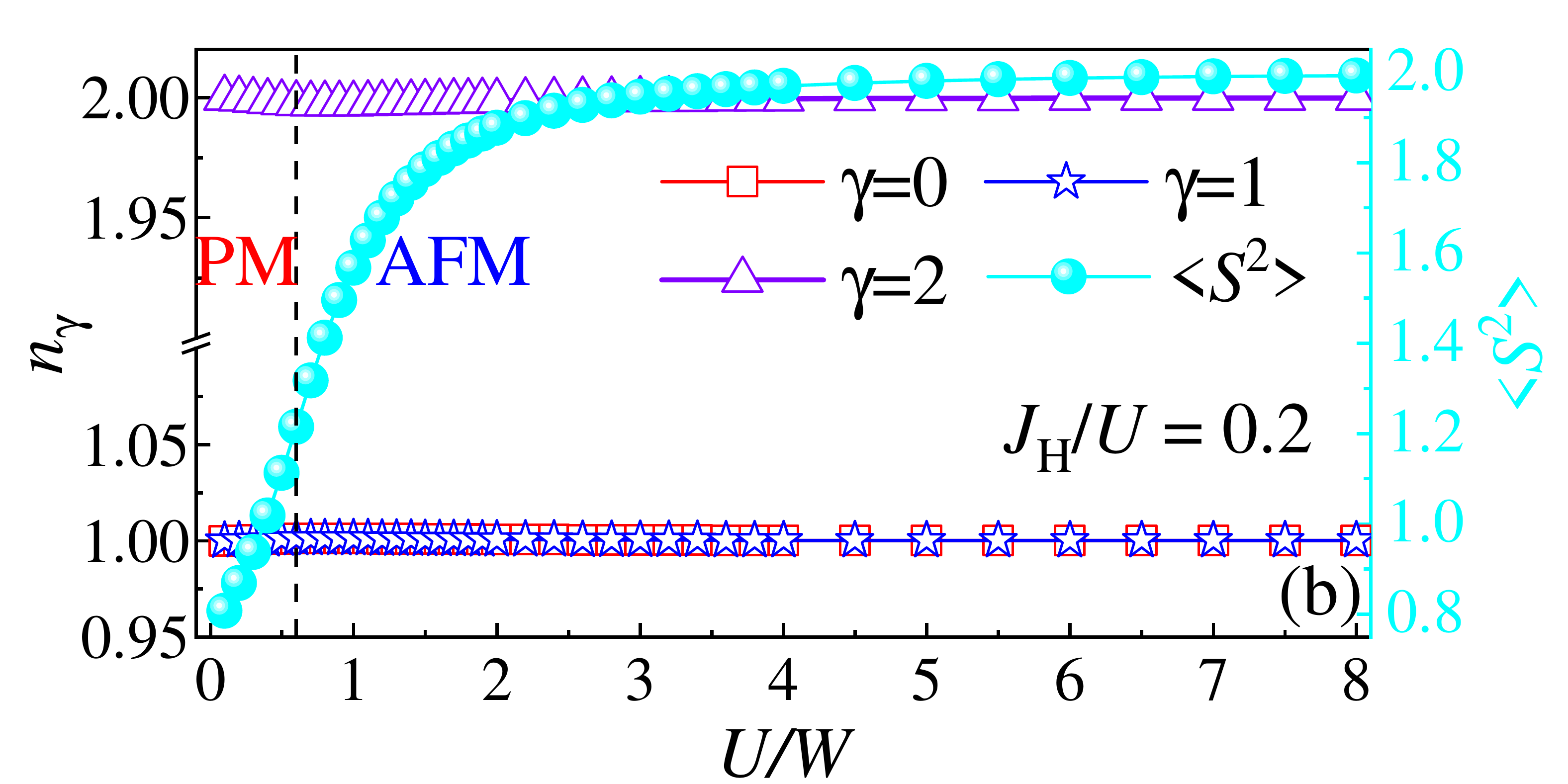}
\caption{Orbital-resolved occupation number $n_{\gamma}$, averaged total spin-squared $\langle{{S}}^2\rangle$ vs. $U/W$, at (a) $J_{H}/U = 0.15$ and (b) $J_{H}/U = 0.2$, respectively. We used a $24$-site chain with NN hoppings for four electrons in three orbitals.}
\label{Fig12}
\end{figure}

In addition, we also calculated the site-average occupancy of different orbitals $n_{\gamma}$ vs $U/W$, for $J_H/U = 0.15$ and $J_{H}/U = 0.2$, respectively. As shown in Fig.~\ref{Fig12}, the population of orbital $\gamma = 2$ is $2$ for the whole region of $U/W$, and this orbital decouples from the system. Furthermore, the other two orbitals $\gamma = 0$ and $\gamma = 1$ remain half-filled for all values of $U/W$.  At the intermediate electronic correlation region, we observed a stable AFM Mott-insulating behavior, different from our previous studies on the same multi-orbital model (four electrons in three orbitals) when  using different hopping matrix elements corresponding to other materials, such as when we reported an orbital-selective Mott phase~\cite{zhang:2021}, FM insulating state~\cite{Lin:prl21}, block AFM phase~\cite{Herbrych:osmp1} and a non-collinear spiral phase~\cite{Herbrych:pnas}. In this case, the system is in a spin $S=1$ per site Mott-insulator staggered AFM state. Thus, increasing $U/W$ opens a gap. Furthermore, the average value of the squared local spin averaged over all the sites $\langle{{S}}^2\rangle$ is also displayed in Fig.~\ref{Fig12}, as a function of $U/W$. With increasing $U/W$, as the system becomes Mott insulating and antiferromagnetically ordered, $\langle S^2\rangle$ saturates to a value of 2, as expected. Note that while the limit of large $U$ may be considered naively as always leading to an AFM state, our study shows that staggered ordered develops at intermediate coupling already, which requires a special calculation as shown here. Moreover, AFM order may be obvious at a density of one electron per orbital, but in our case we have four electrons in three orbitals. The many publications cited before show that when these two numbers are not equal, the magnetic order can be of a different class. Thus, by no means it is obvious {\it a priori} that AFM would develop in our system.

\section{V. Conclusions}

In this publication, we have systematically studied the compounds $M$OCl$_2$ ($M$ = Ru or Os) by using first-principles DFT and also DMRG calculations. In this system with $d^4$ electronic configurations, the ferroelectric distortion and  Peierls instabilities disappear, leading to an undistorted I${\rm mmm}$ phase. Furthermore, with {\it ab initio} DFT calculations, we observed a strongly anisotropic electronic structure along the $a$-axis. Based on the Wannier functions from first-principles calculations, we calculated the relevant hopping amplitudes and crystal-field splitting energies of the $t_{2g}$ orbitals for the Os atoms.  In this case, this system is in a $S=1$ state, instead of a $J = 0$ singlet groundstate, due to the large crystal-field splitting energy (between $d_{xz/yz}$ and $d_{xy}$ orbitals) and large nearest-neighbor hopping. In addition, based on DFT calculations, we also found strongly anisotropic magnetic structures with strong coupling along the $a$-axis and weak coupling along the $b$-axis for both RuOCl$_2$ and OsOCl$_2$. In this case, the coupling along the $M$-O chain leads to staggered magnetic order with $\pi$ wavevector, and the coupling along the $M$-Cl chain direction is weak. Hence, as expressed before, remarkably these systems can be regarded as ``effective 1D'' systems.

In addition, we constructed a multi-orbital Hubbard model for the $M$-O chains. The staggered AFM with $\uparrow$-$\downarrow$-$\uparrow$-$\downarrow$ order was found to be dominant in our DMRG calculations, in agreement with DFT calculations.
Different from the previously well-studied oxide dichlorides $M$O$X_2$ ($M$ = V, Ta, Nb, Ru and Os; $X$ = halogen element) with $d^1$ and $d^2$ configurations, note that {\it thus far almost no research has been reported for other electronic densities $n$ of the $M$ atoms in this family}.
Thus, we believe our results for $M$OCl$_2$ ($M$ = Ru or Os) will provide guidance to experimentalists and theorists working in the
oxide dichlorides family at the novel density $n$ studied here.

\section{Acknowledgments}
The work of Y.Z., L.-F.L., A.M., T.A.M., and E.D. was supported by the U.S. Department of Energy (DOE), Office of Science, Basic Energy Sciences (BES), Materials Sciences and Engineering Division. G.A. was partially supported by the Scientific Discovery through Advanced Computing (SciDAC) program funded by U.S. DOE, Office of Science, Advanced Scientific Computing Research and BES, Division of Materials Sciences and Engineering.

\section{APPENDIX}

\subsection{A. Wannier functions}

According to the crystal-splitting analysis and electronic structures in the previous section, the $e_g$ orbitals of Ru or Os ($d_{x^2-y^2}$ and $d_{3z^2-r^2}$) are located at high energy on the conduction band, far away from the Fermi level with a large energy splitting between $e_g$ and $t_{2g}$ orbitals. In this case, those systems can be regarded as four electrons on a three $t_{2g}$ orbitals low-energy model. To better understand those low-energy orbitals, we constructed three disentangled Wannier functions based on the MLWFs method~\cite{Mostofi:cpc}, involving the $t_{2g}$ orbital basis $d_{xy}$, $d_{yz}$, and $d_{xz}$ for each Ru or Os atom in the NM phase without SOC.

\begin{figure}
\centering
\includegraphics[width=0.48\textwidth]{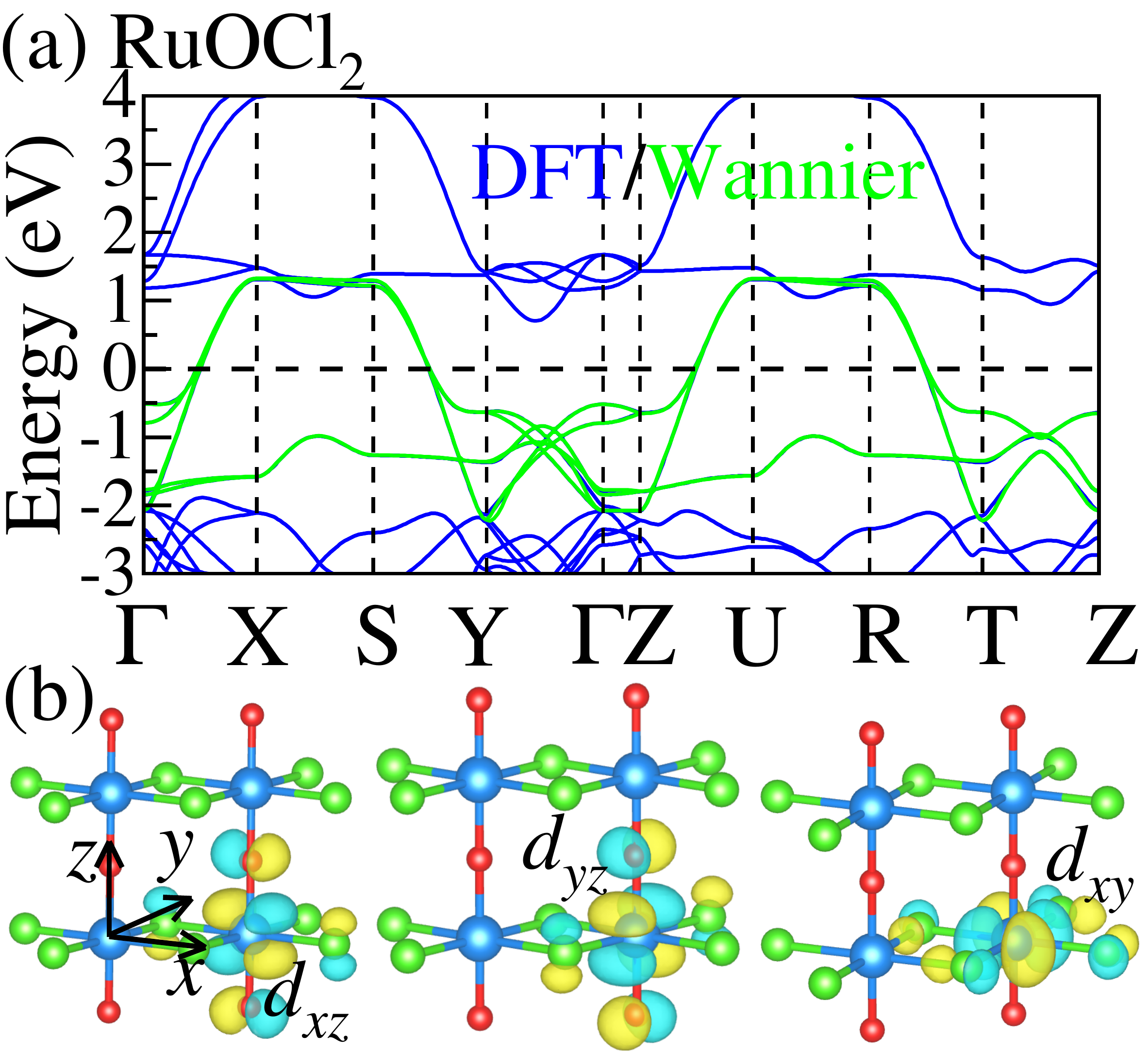}
\includegraphics[width=0.48\textwidth]{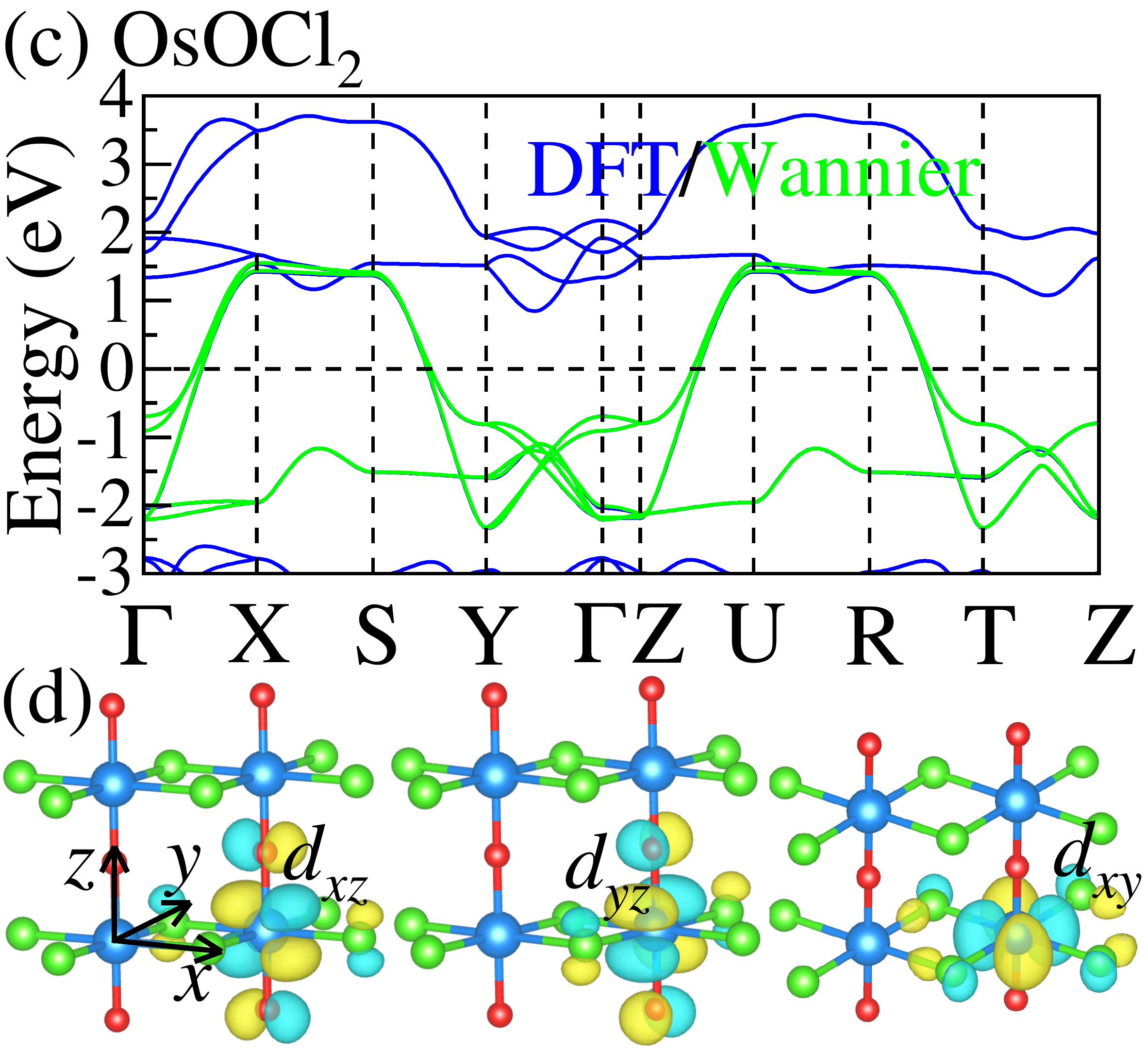}
\caption{ (a) and (c) DFT and Wannier bands of the conventional cell of RuOCl$_2$ and OsOCl$_2$, respectively. The Fermi level is shown with dashed horizontal lines. Note that near the Fermi level the blue DFT bands are totally hidden by the green Wannier bands, indicating the high quality of the fit.  (b) and (d) are Wannier functions of the three Ru or Os $t_{2g}$ orbitals, with lobes of opposite signs colored as blue and yellow. $M$ ($M$ = Ru or Os), O and Cl atoms are in blue, red and green, respectively. The local basis are marked in the inset of (b) and (d), with the $x$- or $y$-axis along the $M$-Cl directions, while the $z$-axis is along the $a$-axis.}
\label{Fig13}
\end{figure}

Figures~\ref{Fig13} (a) and (c) indicate that DFT bands are fitted very well with the Wannier bands obtained from MLWFs. In fact, the blue DFT bands near the Fermi level are totally hidden by the green Wannier bands, because of the quality of the fit. As displayed in Figs.~\ref{Fig13} (b) and (d), those orbitals obtained from MLWFs clearly display $d_{xz}$, $d_{yz}$ and $d_{xy}$ characteristics. Based on the Wannier function basis \{$d_{xz}$, $d_{yz}$, $d_{xy}$\}, here referred to as $\gamma$ =  \{0, 1, 2\}, we deduced the on-site energies of the three $t_{2g}$ orbitals, as well as the hopping parameters, for RuOCl$_2$ and OsOCl$_2$, respectively.

First, we obtained the on-site matrices for the Ru or Os atoms, using the basis \{$d_{xz}$, $d_{yz}$, $d_{xy}$\}:
\begin{equation}
\begin{split}
t_{onsite}^{Ru} =
\begin{bmatrix}
           d_{xz}      &       d_{yz}   &     d_{xy}   \\
           3.901	   &       0.000	    &      0.000	\\
           0.000	   &       3.903	    &      0.000	\\
           0.000	   &       0.000	    &      2.680	
\end{bmatrix},\\
\end{split}
\label{onsite1}
\end{equation}

\begin{equation}
\begin{split}
t_{onsite}^{Os} =
\begin{bmatrix}
           3.721	   &       0.000	    &      0.000	\\
           0.000	   &       3.727	    &      0.000	\\
           0.000	   &       0.000	    &      2.255	
\end{bmatrix}.\\
\end{split}
\label{onsite1}
\end{equation}

Furthermore, we also obtained the NN hopping matrices along the $a$-axis.

For RuOCl$_2$:
\begin{equation}
\begin{split}
t_{\gamma\gamma'}^a =
\begin{bmatrix}
         -0.632     &   0.009  &   0.000	   	       \\
          0.009     &  -0.633  &   0.000	   	       \\
          0.000	    &   0.000  &  -0.012	
\end{bmatrix}.\\
\end{split}
\end{equation}

For OsOCl$_2$:
\begin{equation}
\begin{split}
t_{\gamma\gamma'}^a =
\begin{bmatrix}
         -0.713     &   0.013  &   0.000	   	       \\
          0.013     &  -0.717  &   0.000	   	       \\
          0.000	    &   0.000  &  -0.011	
\end{bmatrix}.\\
\end{split}
\end{equation}

In addition, we also obtained the nearest-neighbor hopping matrices along the $b$-axis.

For RuOCl$_2$:
\begin{equation}
\begin{split}
t_{\gamma\gamma'}^b =
\begin{bmatrix}
          0.003     &  -0.081  &   0.000	   	       \\
         -0.081     &   0.020  &   0.000	   	       \\
          0.000	    &   0.000  &  -0.070	
\end{bmatrix}.\\
\end{split}
\end{equation}

For OsOCl$_2$:
\begin{equation}
\begin{split}
t_{\gamma\gamma'}^b =
\begin{bmatrix}
          0.004     &  -0.103  &   0.000	   	       \\
         -0.103     &   0.033  &   0.000	   	       \\
          0.000	    &   0.000  &  -0.103	
\end{bmatrix}.\\
\end{split}
\end{equation}

All the on-site and hopping matrix elements are in eV units. Note that the angle formed by $M$-Cl-$M$ ($M$ = Ru or Os) is not $90 ^{\circ}$, causing a slight deviation of the local $y$-axis from the direction of the $M$-Cl bond, as shown in Fig.~\ref{Fig1}. Hence, there are tiny differences in the values of the on-site energies and hopping amplitudes between the otherwise degenerate $d_{xz}$ and $d_{yz}$ orbitals. Furthermore, the NN hoppings between each Ru or Os layer along the $c$-axis are quite small and can be neglected compared with the others. In addition, the NN hopping along the $d_{xz/yz}$ $a$-axis are much larger than the NN hoppings along the $b$-axis for both RuOCl$_2$ and OsOCl$_2$. Then, the magnetic properties of $M$OCl$_2$ ($M$ = Ru or Os) are mainly determined by the hopping along the $a$-axis between NN Ru-Ru or Os-Os atoms, leading to strong anisotropic magnetism. Based on the NN hopping matrices along the $a$-axis of RuOCl$_2$ and OsOCl$_2$, the hopping of the diagonal elements of $d_{xz/yz}$ orbitals are dominant, leading to a strong AFM coupling.

\subsection{B. Band structures of NM states with $U$}
\begin{figure}
\centering
\includegraphics[width=0.48\textwidth]{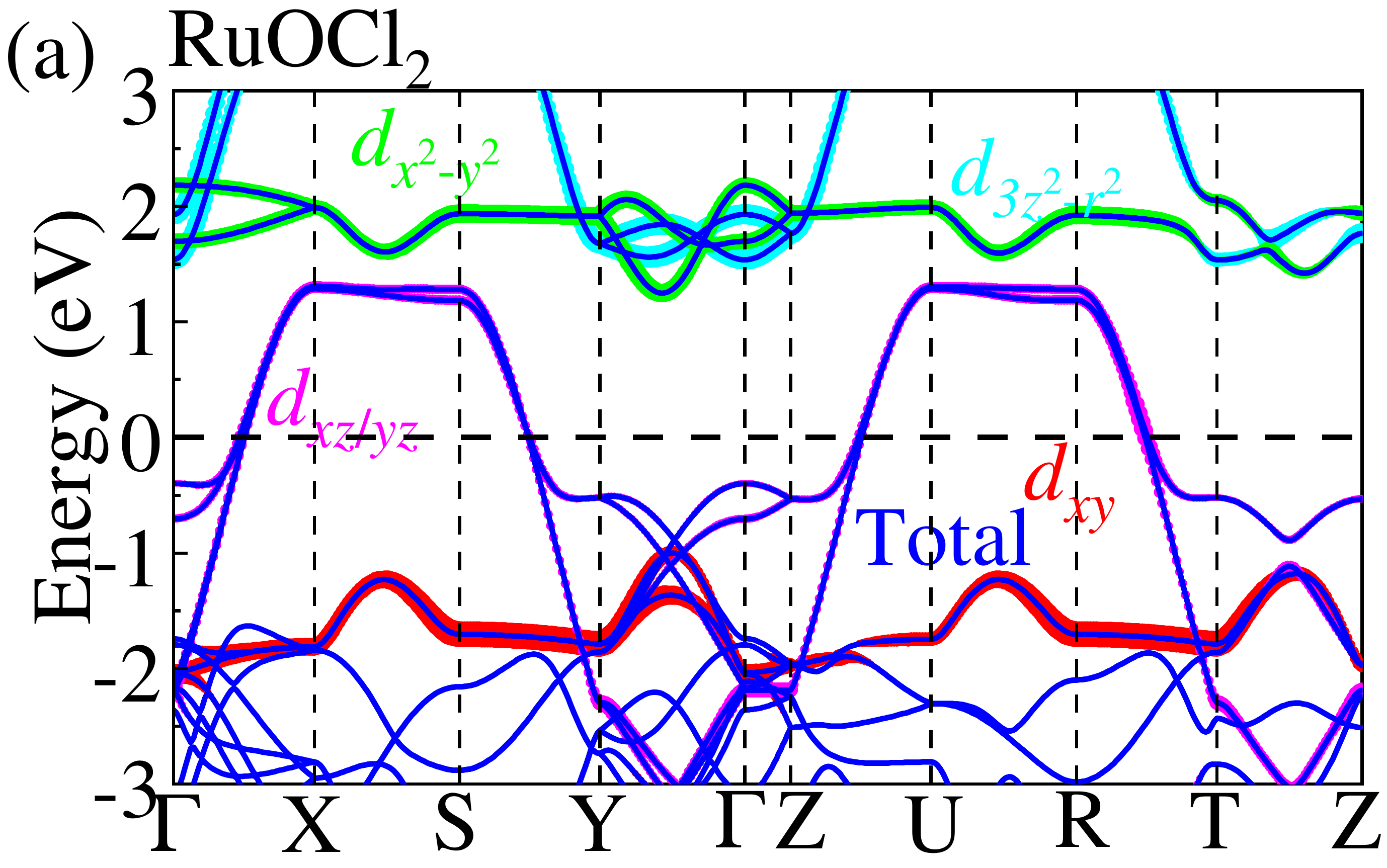}
\includegraphics[width=0.48\textwidth]{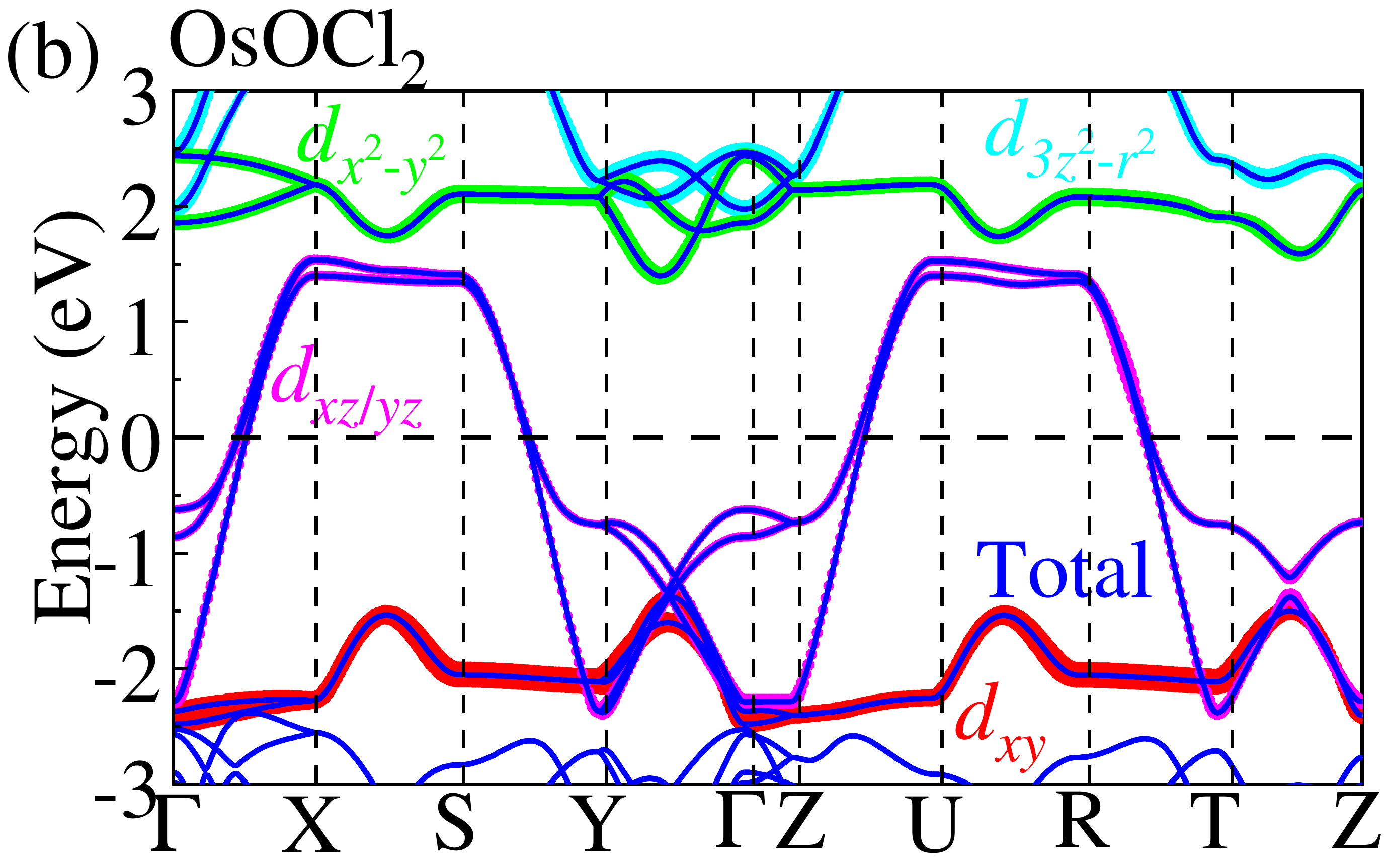}
\caption{Projected band structures of the undistorted I${\rm mmm}$ NM state. (a) $U = 3$ eV and $J = 0.6$ eV for RuOCl$_2$. (b) $U = 2$ eV and $J = 0.4$ eV for OsOCl$_2$, respectively. Note that the local \{$x$, $y$, $z$\} axes of projected orbitals are marked in
Fig.~\ref{Fig1}, where the $z$-axis is the $a$-axis and $x$ or $y$ axes are along the $M$-Cl directions. The weight of each Ru or Os orbital is represented by the size of the (barely visible) circles.}
\label{Fig14}
\end{figure}

Next, let us discuss the effect of different values of $U$ on the nonmagnetic  state of RuOCl$_2$ and OsOCl$_2$. As shown in Fig.~\ref{Fig14}, the band structures of the NM states are almost unchanged. Hence, the crystal-field splitting, and nearest-neighboring hopping, do not change much.

\subsection{C. Band structures of our LSDA calculations}
\begin{figure}
\centering
\includegraphics[width=0.48\textwidth]{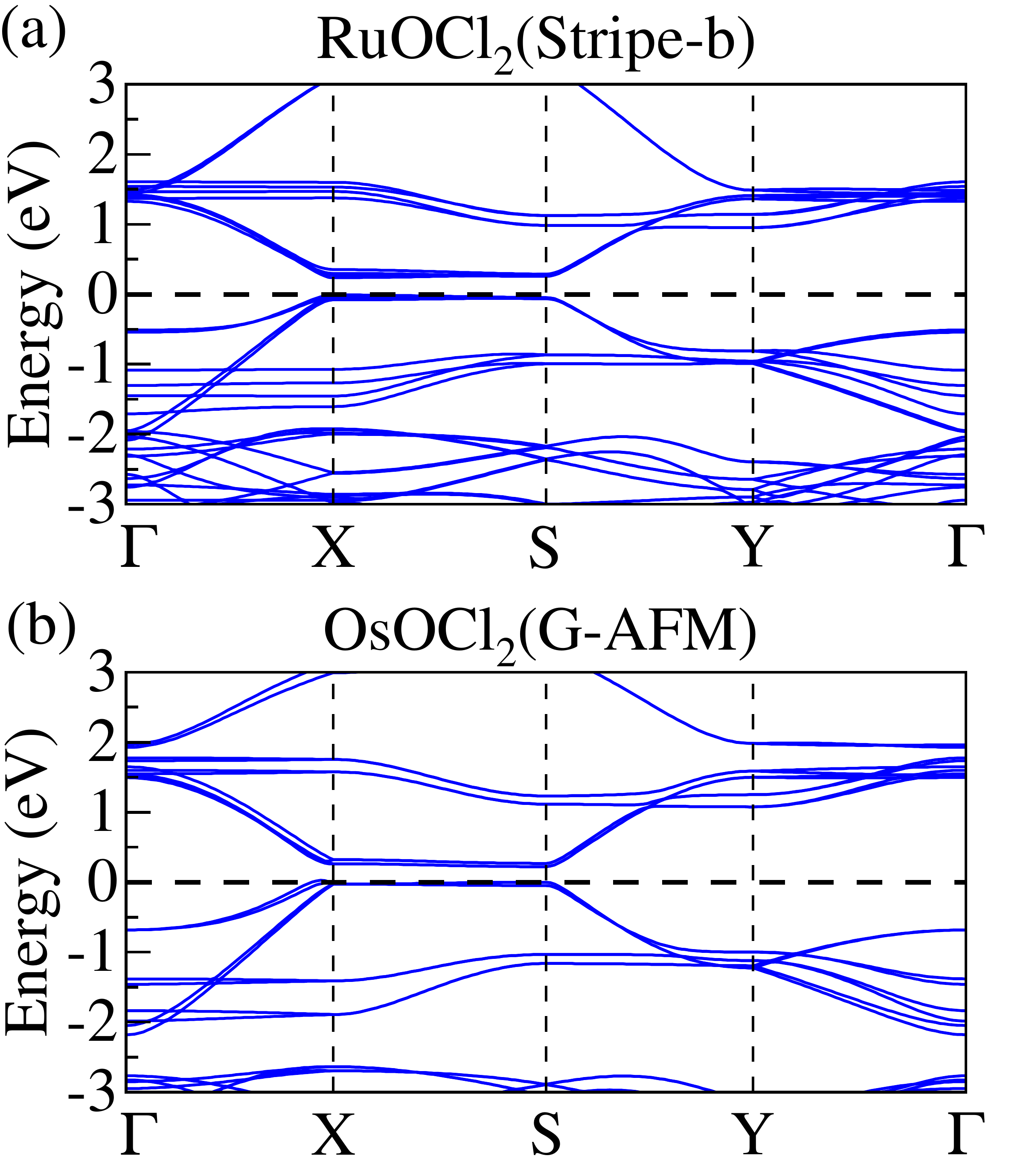}
\caption{Band structures of LSDA calculations for (a) the Stripe-b AFM state of RuOCl$_2$ and (b) the G-AFM state of OsOCl$_2$ without SOC, respectively. The Fermi level is shown with dashed horizontal lines.}
\label{Fig15}
\end{figure}

Using pure LSDA calculations, we also obtained the Stripe-b and G-type AFM insulating ground states for RuOCl$_2$ and OsOCl$_2$, respectively. For the benefit of the readers, the band structures arising from LSDA calculations of the ground state of  RuOCl$_2$ and OsOcl$_2$  are displayed in Fig.~\ref{Fig15}.

 \end{document}